\documentclass[fleqn,usenatbib]{mnras}
\usepackage{newtxtext,newtxmath}
\usepackage[T1]{fontenc}

\DeclareRobustCommand{\VAN}[3]{#2}
\let\VANthebibliography\thebibliography
\def\thebibliography{\DeclareRobustCommand{\VAN}[3]{##3}\VANthebibliography}

\usepackage{graphicx}	% Including figure files
\usepackage{amsmath}	% Advanced maths commands
\newcommand{\enzo}{{\small ENZO}}

\title[Magnetic field evolution  in cosmic filaments] {Magnetic field evolution  in cosmic filaments with LOFAR data}

\author[E.~Carretti et al.]{
E.~Carretti,$^{1}$\thanks{E-mail: carretti@ira.inaf.it (EC)}
S.~P.~O'Sullivan,$^{2}$
V.~Vacca,$^{3}$ 
F.~Vazza,$^{4,1,6}$
C.~Gheller,$^{1}$
T.~Vernstrom,$^{5}$
A.~Bonafede$^{4,1}$ \\
\\
$^{1}$INAF, Istituto di Radioastronomia, Via Gobetti 101, 40129 Bologna, Italy\\
$^{2}$School of Physical Sciences and Centre for Astrophysics \& Relativity, Dublin City University, Glasnevin D09 W6Y4, Ireland\\
$^{3}$INAF, Osservatorio Astronomico di Cagliari, Via della Scienza 5, 09047 Selargius (CA), Italy\\
$^{4}$Dipartimento di Fisica e Astronomia, Universit\'a di Bologna, via Gobetti 93/2, 40122 Bologna, Italy\\
$^{5}$ICRAR, The University of Western Australia, 35 Stirling Hw, 6009 Crawley, Australia  \\
$^{6}$Hamburger Sternwarte, University of Hamburg, Gojenbergsweg 112, 21029 Hamburg, Germany\\
}

\date{Accepted XXX. Received YYY; in original form ZZZ}

\pubyear{2022}

\begin{document}
\label{firstpage}
\pagerange{\pageref{firstpage}--\pageref{lastpage}}
\maketitle

\begin{abstract}
Measuring the magnetic field in cosmic filaments reveals how the Universe is magnetised and the process that magnetised it. Using the  Rotation Measures (RM)  at 144-MHz  from the LoTSS DR2 data, we analyse the  rms of the RM extragalactic component as a function of redshift to investigate the evolution with redshift of the magnetic field in  filaments. From previous  results, we find that the extragalactic term of the RM rms at 144-MHz is  dominated by the contribution from filaments (more than 90 percent). Including  an error term  to account for the minor contribution local to the sources, we fit the data with a model of the physical filament magnetic field, evolving as $B_f = B_{f,0}\,(1+z)^\alpha$ and with a density  drawn from  cosmological simulations of five  magnetogenesis scenarios. We find that the best-fit slope is in the range $\alpha = [-0.2, 0.1]$ with uncertainty of $\sigma_\alpha = 0.4$--0.5, which is consistent with no evolution. The comoving field  decreases with redshift with a slope of $\gamma = \alpha - 2 = [-2.2, -1.9]$. The mean field strength at $z=0$ is in the range  $B_{f,0}=39$--84~nG. For a typical filament gas overdensity of $\delta_g=10$ the filament field strength at $z=0$ is in the range  $B_{f,0}^{10}=8$--26~nG. A primordial stochastic magnetic field model with  initial comoving field of $B_{\rm Mpc} = 0.04$--0.11~nG is favoured.   The  primordial uniform field model is rejected. 
\end{abstract}


%%%%%%%%%%%%%%%%%%%%%%%%%%%%%%%%%%%%%%%%%%%%%%%%%%

%%%%%%%%%%%%%%%%% BODY OF PAPER %%%%%%%%%%%%%%%%%%

%----
\section{Introduction}
\label{sec:intro}
%----
The evolution with cosmic time of the magnetic field is essential to understand how the present Universe is magnetised and the process of magnetogenesis \citep[e.g.,][]{
%2011ApJ...738..134A, 2015A&A...580A.119V, 
2016RPPh...79g6901S, 
%2017CQGra..34w4001V, 
%2020MNRAS.495.2607O, 2021Galax...9..109V, 
2021MNRAS.505.5038A, 2021MNRAS.500.5350V}.
%, 2022ApJ...929..127M}. 
Cosmic web filaments are a sweet spot for this, for they are  not yet as processed by cosmic  evolution as   galaxy clusters are, thus preserving the signature of the initial magnetogenesis scenario \citep[e.g.,][]{2017CQGra..34w4001V, 2021Galax...9..109V, 2021MNRAS.500.5350V, 2022ApJ...929..127M}, while also  possessing stronger fields than in voids, which makes their detection easier. Magnetogenesis scenarios can be broadly subdivided into primordial, where the field is generated either during Inflation or in some early phase-transition before the recombination \citep[e.g.][]{1988PhRvD..37.2743T, 1994RPPh...57..325K, 2019JCAP...11..028P, 2022MNRAS.515..256P}, and late, where the field is generated at low redshift by dynamo amplification or  astrophysical sources that inject it in the intergalatic medium (IGM) as magnetic bubbles \citep[e.g.,][]{1994RPPh...57..325K, 2006MNRAS.370..319B, 2017CQGra..34w4001V}.

The Rotation Measure (RM) of extragalactic sources measures the  magnetic field component along the line-of-sight weighted by the free-electron number density and integrated along the entire line-of-sight. It is a powerful tool to investigate magnetic field properties of the Galaxy \citep[e.g.,][]{2012ApJ...757...14J, 2022arXiv220910819D}, the environment local to the source \citep[e.g.,][]{2008ApJ...676...70K}, or the intervening IGM between the source and the observer \citep[e.g.,][]{2019ApJ...878...92V, 2020MNRAS.495.2607O}. 

%The  evolution of RM with redshift was investigated in several papers  in  past decades, aimed at understanding   the evolution of the magnetic field with cosmic time in different environments, with no clear detection of an IGM signature \citep[e.g., see][and references therein]{2022MNRAS.512..945C}}, until   the work  by \citet{2022MNRAS.512..945C} who, using some 1,000 RMs  measured at  low frequency with  LOFAR (O'Sullivan et al.  submitted),  found  that an IGM origin of  the RMs  is favoured at those frequencies.   

The detection of the  radio emission of cosmic filaments and  of their magnetic field through synchrotron emission and RM was the subject of intense research in the past few years. 
Upper limits were found with different approaches: cross-correlating large radio maps with the large-scale  galaxy distribution  \citep{2017MNRAS.467.4914V, 2017MNRAS.468.4246B};  analysing   RMs of giant radio  galaxies \citep{2019A&A...622A..16O, 2020A&A...638A..48S}; cross-correlating RMs with the galaxy distribution \citep{2021MNRAS.503.2913A};  simulations constrained by observations or non-detections \citep{2018MNRAS.479..776V, 2021A&A...652A..80L}.   
Intracluster bridges of radio emission were detected in a few galaxy clusters\footnote{\citet{1989Natur.341..720K} proposed that the structure they found, stretching out of the Coma galaxy cluster halo, was an intercluster bridge connecting the Coma cluster to the cluster A1367. Later observations have shown it is an intracluster bridge in the Coma cluster connecting the halo to the SW relic \citep[e.g.,][]{2011MNRAS.412....2B,2022ApJ...933..218B}. }  \citep[e.g.,][]{1989Natur.341..720K,2011MNRAS.412....2B, 2022ApJ...933..218B,2022A&A...659A.146D}. 
A detection of the synchrotron emission from an intercluster bridge connecting close pairs of merging clusters was obtained by \citet{2019Sci...364..981G} \citep[see also][]{2020MNRAS.499L..11B}, establishing the presence of magnetic fields in the IGM beyond cluster outskirts. \citet{2021MNRAS.505.4178V} and \citet{2022MNRAS.512..945C} made a further step ahead, first detecting fields of the general, weaker  filaments of the cosmic web of 30--60~nG and $\approx$30~nG, through stacking of synchrotron emission and measuring the RM evolution with redshift, respectively.

The evolution with redshift of the RM and  average magnetic field of the Universe were investigated by several authors (e.g., \citealt{2014MNRAS.442.3329X} and references therein and in \citealt{2022MNRAS.512..945C}), but hampered by the separation of local and IGM components.
\citet{2022MNRAS.515..256P}  separated the IGM term using  the differential RM of close pairs of galaxies from the same, low frequency RM catalogue we used in paper I, and measured  the evolution with redshift of the average magnetic field of the Universe.

In \citet{2022MNRAS.512..945C}, hereafter Paper~I, we used the RM catalogue  at 144-MHz (O'Sullivan et al. submitted)  derived from LoTSS DR2 (LOFAR Two-metre Sky Survey Data Release 2, \citealt{2017A&A...598A.104S, 2019A&A...622A...1S, 2022A&A...659A...1S}) data to  measure the behaviour of RM  in redshift bins out to $z=2$, after subtracting off the Galactic contribution, and the behaviour versus the fractional polarization $p$. We found that the former  is consistent with no evolution, and  the latter is flat with $p$. After a comparison with the RM and $p$ measured at 1.4-GHz of the same sources, we found that an IGM origin of the RMs is favoured and estimated a magnetic field in filaments of $\approx$30~nG, as reported above. We assumed no evolution for magnetic field and electron number density, however, except assuming the mean electron number density at $z=0.7$.

%\citet{2022MNRAS.515..256P} estimated the evolution with redshift of the average magnetic field of the Universe out to $z=2$  for the first time, using  the differential RM of close pairs of galaxies from the same RM catalogue we used in paper I. 

This work is a follow-up of Paper~I, aimed at investigating the evolution  with redshift of the  magnetic field in cosmic filaments, adding in the evolution of the quantities involved. It is conducted within   the Magnetism Key Science Project (MKSP) of LOFAR and uses the RM catalogue at 144-MHz employed in Paper I  (O'Sullivan et al. submitted), which is derived from  LoTSS DR2  \citep{2022A&A...659A...1S} Stokes $Q$ and $U$ data cubes in a collaborative effort between the LOFAR Surveys Key  Science Project\footnote{https://lofar-surveys.org/} and the MKSP. It also uses dedicated cosmological magneto-hydrodynamical (MHD) simulations of a set of magnetogenesis scenarios,  to draw realistic density distributions from and to compare our results with. We find the RM evolution with redshift a  powerful way to discriminate  between cosmological magnetogenesis models. 

This paper is organised as follows.  Section~\ref{sec:data} describes the RM data and the RM rms  in redshift bins out to $z=3$, after subtracting off the Galactic contribution. Section~\ref{sec:simul} describes the MHD simulations of the magnetogenesis scenarios we used for this work. Section~\ref{sec:zevo} contains  our analysis of the evolution with redshift of the magnetic field in cosmic filaments, including best-fits to the data, considerations on the environment where the low-frequency RMs are produced, and estimates of the predictions of the magnetogenesis scenarios we considered. Finally, Sections~\ref{sec:discussion} and~\ref{sec:conc} present our discussion and conclusions, with a comparison of our results with  the magnetogenesis scenarios we considered. 

Throughout the paper we assume the flat $\Lambda$CDM cosmological model assumed in the simulations of Section~\ref{sec:simul}, with $H_0 = 67.8$~km~s$^{-1}$~Mpc$^{-1}$, $\Omega_M = 0.308$, $\Omega_\Lambda =0.692$,  $\Omega_b = 0.0468$, and $\sigma_8=0.815$ \citep{2016A&A...594A..13P}. Errors  refer to 1-sigma uncertainties. 
%We also use the term $h=H_0 /100$~km~s$^{-1}$~Mpc$^{-1}$.

%----
\section{RM data}
\label{sec:data}
%----
%
\subsection{LoTSS DR2 RM catalogue}
This work   is based on the RM catalogue derived from the   LoTSS DR2 survey using its Stokes $Q$ and $U$ data cubes   (O'Sullivan et al. submitted). Here we report the main catalogue features relevant to this work and refer to the description paper for full details.   It consists of 2,461 RMs detected, over 5,720-deg$^2$, in the frequency range 120--168~MHz with channels of width of 97.6-kHz, and angular resolution of 20-arcsec. RMs were  obtained using  RM-synthesis \citep{1966MNRAS.133...67B, 2005A&A...441.1217B}. The RM error budget  is dominated by ionospheric RM correction residuals that  can be as large as 0.1--0.3 rad m$^{-2}$  \citep{2013A&A...552A..58S,2019MNRAS.483.4100P}.
In this dataset it is estimated to be $\approx 0.05$-rad~m$^{-2}$ (O'Sullivan et al. submitted). 
A total number of 1,949 sources had a positive cross-match with redshift catalogues, 1,046 of which are spectroscopic redshifts. 

We did not use photometric redshifts of the identified sources because of their  median error of $\sigma_{z,\rm  phot} \approx 0.1$, comparable to or larger than the redshift bin width used here, and kept  sources with spectroscopic redshift only.   A Galactic cut of  $|b|  >  25^\circ$ was  applied to exclude the region with highest Galactic RM values. 
The median redshift is $\approx 0.5$ and only a handful of sources have redshift $z > 3$ (see Figure 1 of Paper I  for the redshift distribution).  We limited our  analysis to $z < 3$, which gave our final sample of 1,014 objects.

%
%----
\subsection{Behaviour of RM dispersion} 
\label{sec:rmrms}
%----

%%% Figure %%%%
   \begin{figure*}
   \centering
    \includegraphics[width=\textwidth]{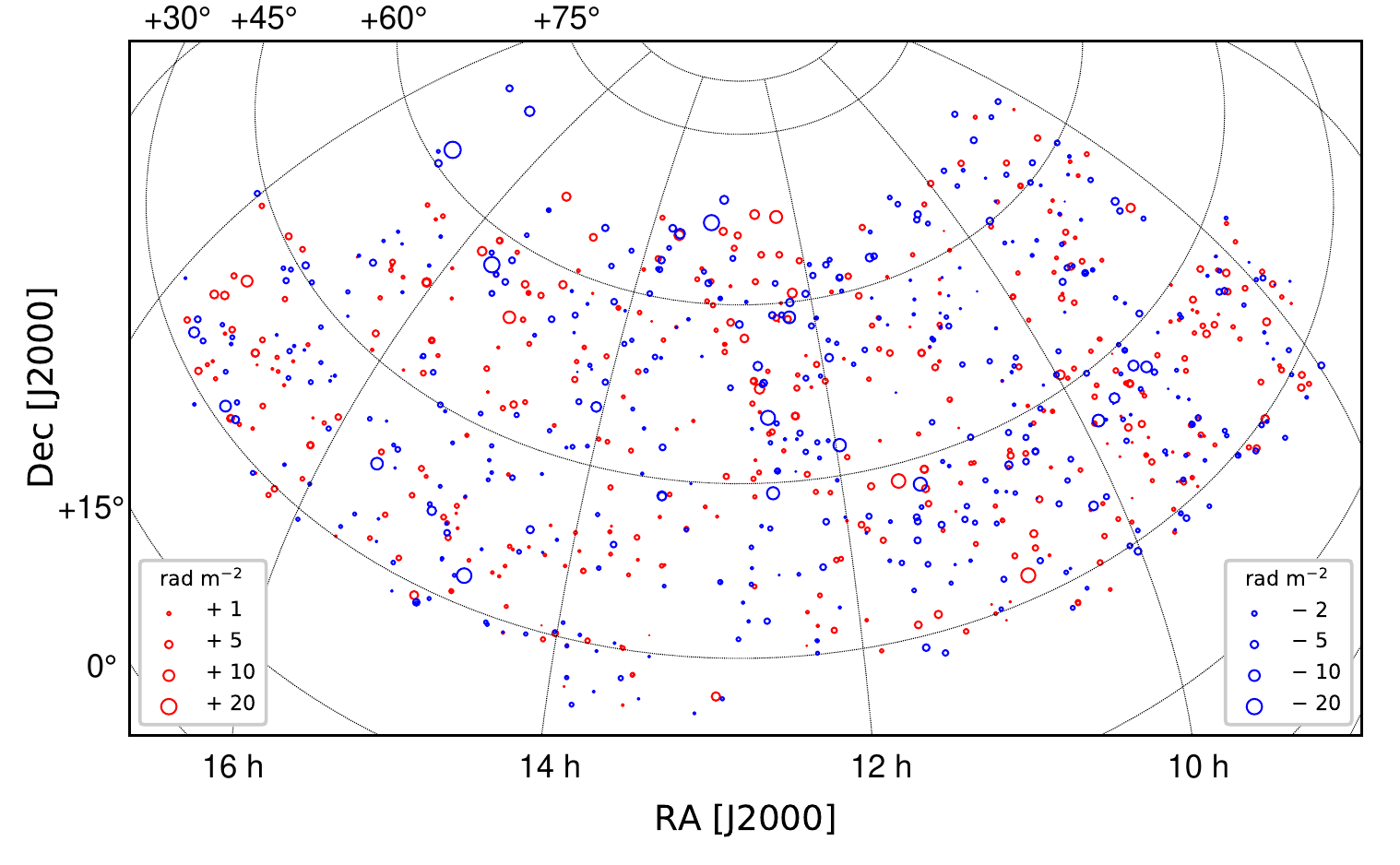}
   \caption{Sky distribution of the RRM sample used in the analysis. The field of the LoTSS DR2 survey centred at $\rm RA = 13h$ is shown.}
              \label{fig:rrm_13h}%
    \end{figure*}
%%%%%
%%% Figure %%%%
   \begin{figure}
   \centering
    \includegraphics[width=\columnwidth]{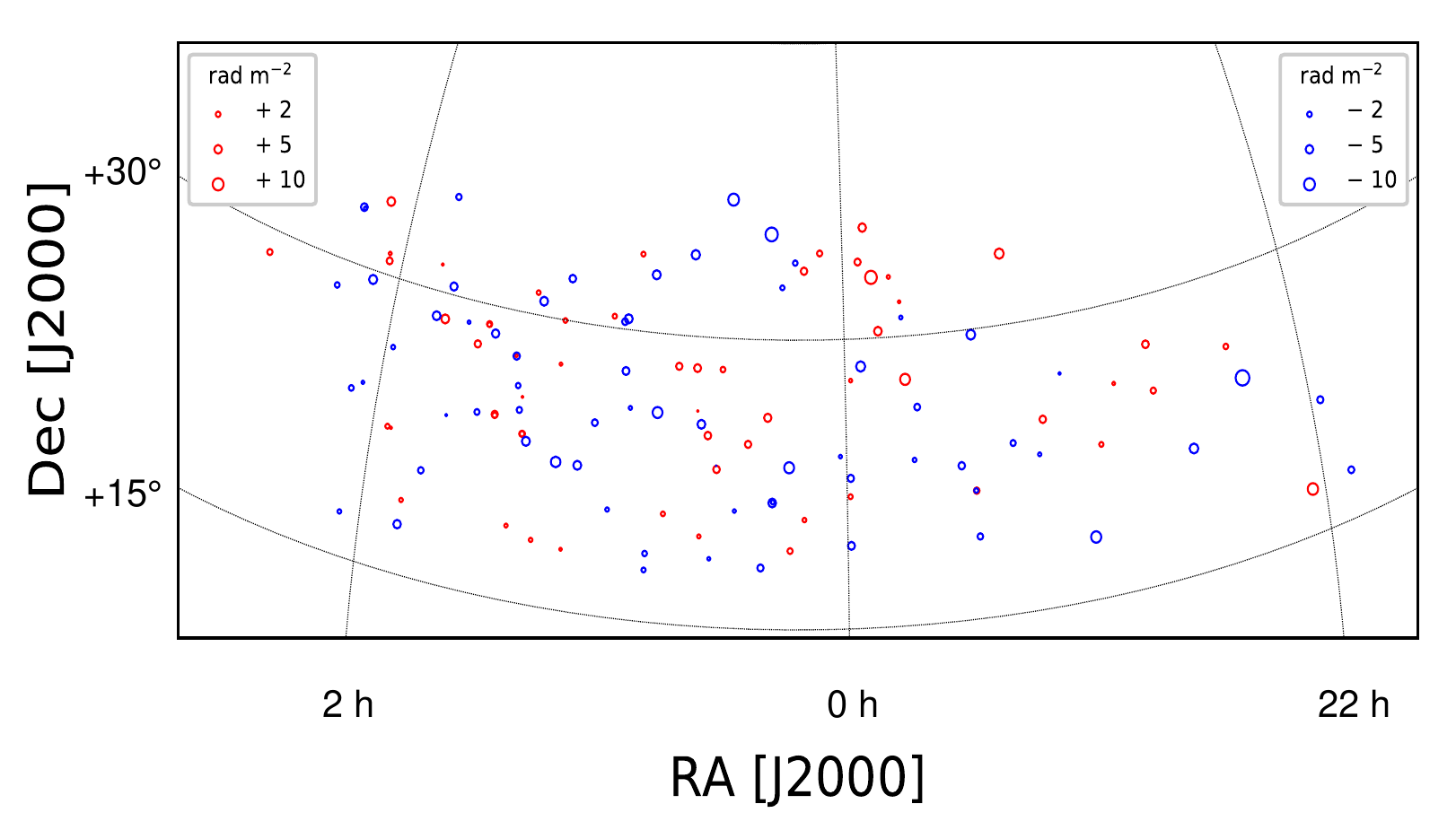}
   \caption{As for Figure \ref{fig:rrm_13h}, except the field  centred at $\rm RA = 0h$ is shown.}
              \label{fig:rrm_0h}%
    \end{figure}
%%%%%
%%% Figure %%%%
   \begin{figure*}
   \centering
    \includegraphics[width=\columnwidth]{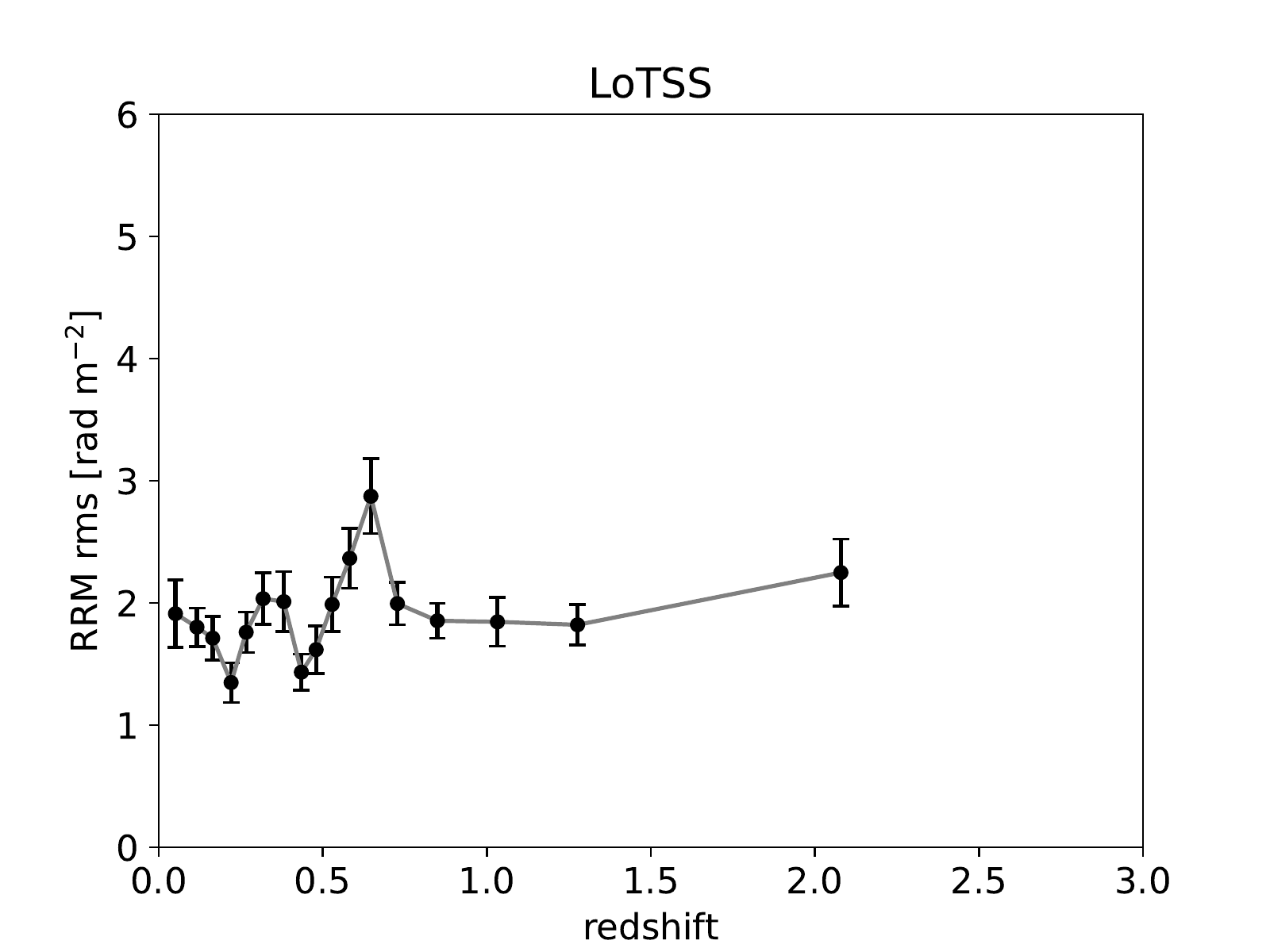}
    \includegraphics[width=\columnwidth]{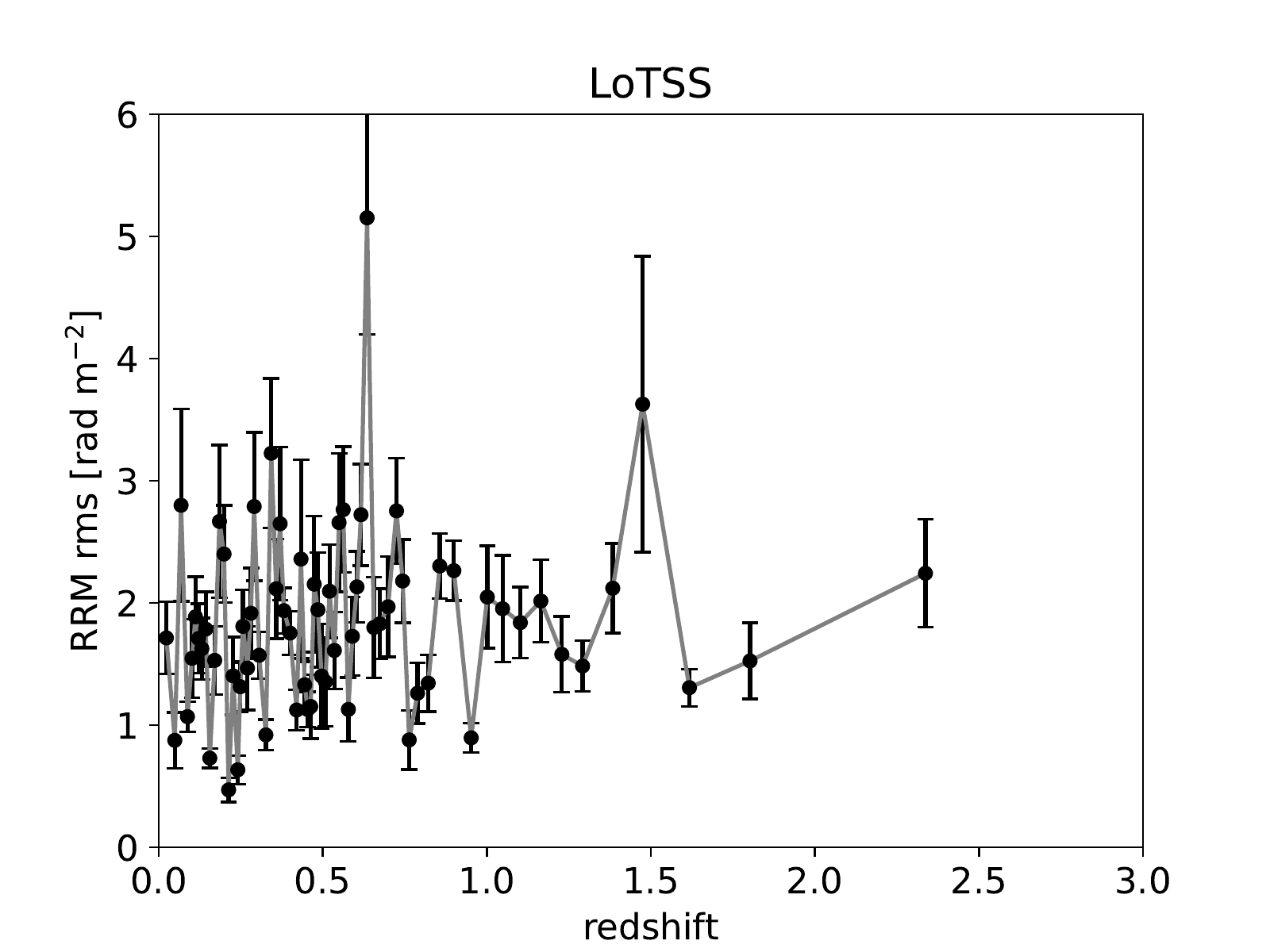}
   \caption{RRM rms in redshift bins with the same number of sources, 60 (left) and 15 (right). }
              \label{fig:rrmvsz}%
    \end{figure*}
%%%%%
The estimate of the evolution with redshift of the RM  extragalactic component is done as for  Paper I, except it is stretched out to $z=3$ and the numbers of bins is increased by $\approx 4$ times. 

The RM of an extragalactic source is a combination of a Galactic component (GRM), an extragalactic term, either local to the source or the IGM intervening between the source and the observer, and the instrumental noise: 
\begin{equation}
    \rm RM = GRM +  RM_{local} + RM_{IGM} + RM_{noise}\, .
\end{equation}
The local term usually is dominated by the environment around the source, such as the intracluster medium of a galaxy cluster \citep[e.g.,][]{2008MNRAS.391..521L}.

The extragalactic component is obtained by subtracting off the Galactic term: 
\begin{equation}
    \rm RRM = RM - GRM
\end{equation}
that  we call the Residual RM (RRM).

Following Paper I, we estimated the GRM at each source position from the Galactic RM map by \citet{2022A&A...657A..43H} as the median of a 1-degree diameter disc centred at the source. We refer to Paper I for  details and motivations. The result is shown in Figures \ref{fig:rrm_13h} and \ref{fig:rrm_0h}, that report the RRMs of the  two fields of the sample.   The GRM error of each source is estimated by bootstrapping, which also captures the GRM variations within the 1-degree disc.

We then computed the  dispersion of the RRM values  $<{\rm RRM}^2>^{1/2}$ (hereafter RRM rms) in redshift bins with the same number of sources per bin, as for Paper I. The quadratic mean of the GRM errors and of the measurement noise {of the sources in each bin} were quadratically subtracted off to remove   their bias\footnote{The square of the measured RRM rms is $<{\rm RRM_{mes}}^2> = <{\rm RRM}^2> + \sigma^2_{\rm GRM} + \sigma^2_{\rm noise}$  where $\sigma^2_{\rm GRM}$ and  $\sigma^2_{\rm noise}$ are the means of the   variances of GRM and measurement noise.}. We excluded outliers, only keeping RRMs  witin 2-sigma, as for Paper I.

The result is shown in  Figure\ref{fig:rrmvsz} for two cases, 60 and 15 sources per bin, that differ in the number of  bins (17 and 68, respectively) and the uncertainty per bin (mean of 0.21 and 0.35~rad~m$^{-2}$). The error is estimated by bootstrapping.
Both cases are consistent with no evolution with redshift, the slope of a linear regression is $0.22\pm 0.17$~rad~m$^{-2}$ and $0.24\pm 0.20$~rad~m$^{-2}$ for the two cases. There is a marginal increase, but at less than 2-sigma significance.  

\section{Cosmological MHD simulations}
\label{sec:simul}
%----

We used the cosmological magneto-hydrodynamical code {\enzo}\footnote{enzo-project.org} to produce  new $\Lambda$CDM  simulations of a volume of $\approx(85\ \mathrm{Mpc})^3$ (comoving) sampled with a static grid of $512^3$ cells, giving a constant spatial resolution of $166$ kpc/cell and a constant mass resolution of  $6.48 \times 10^{8}\ \mathrm{M_{\odot}}$ per dark matter particle. These simulations are qualitatively similar to those analysed in \citet{2017CQGra..34w4001V},  with a few updates, also motivated by the findings of our recent work in \citet[][]{2022MNRAS.515..256P}. Firstly, in this suite of simulations, radiative gas cooling is included in all models, which moderately increases the level of gas clumping in cosmic filaments. Furthermore, we explored additional models of magnetic fields, including an inflationary primordial model following \citet[][]{2021MNRAS.500.5350V}, and a  mixed (astrophysical and primordial) model. Lastly, we produced synthetic lines-of-sight out to a larger redshift  ($z=3$, as opposed to $z=2$ in \citealt{2022MNRAS.515..256P}) using a much larger number of snapshots finely spaced in time, as compared to earlier  work, to monitor evolutionary trends with redshift in a more accurate way.  These simulations  are used to estimate the magnetic field in cosmic filaments instead of the general IGM. Similar to previous projects \citep[e.g.][]{2017CQGra..34w4001V}, we produced different scenarios (five in this case) for the origin and evolution of extragalactic magnetic fields: 

\begin{enumerate}
 \item \textit{``primordial uniform''}:  a primordial uniform volume-filling comoving magnetic field $B_0=0.1\ \mathrm{nG}$ initialised at the beginning of the simulation ($z=40$);
  \item \textit{``primordial stochastic''}:  a tangled primordial magnetic field, with fields scale dependence described by a power law spectrum: $P_B(k) = P_{B0}k^{\alpha_s}$ characterised by a constant spectral index and an amplitude, commonly referred after smoothing the fields within a scale $\lambda=1 \rm ~ Mpc$, using the same approach of \citet[][]{2021MNRAS.500.5350V}. In this work we assumed an initial "blue" spectrum with $\alpha_s=1.0$  and $B_{\rm Mpc}=0.042 \rm ~nG$ (comoving), based on the recent constraints from the combined analysis of the Cosmic Microwave Background with different instruments by \citet{2019JCAP...11..028P}.  We selected this value of $\alpha_s$ from the best constraint provided by previous observational tests \citep{2021Galax...9..109V}. 
  
 \item \textit{``dynamo''}: a uniform initial seed magnetic field of $B_0 = 10^{-11}\ \mathrm{nG}$ (comoving) that can be amplified through 
 "sub-grid" dynamo amplification computed at run-time, which allows the estimation of the hypothetical maximum contribution of a dynamo in low density environments \citep[see][]{2008Sci...320..909R}, where it would be lost due to finite resolution effects \citep[see][for more details]{2017CQGra..34w4001V};
 \item \textit{``astroph''}: a model in which the magnetic field is released in the form of magnetic loops from overdense regions of the simulation, whenever AGN feedback is triggered by local gas overcooling. To maximise the plausible combined effect of star formation driven winds, and AGN feedback, we assumed a large, average of $50 \%$ conversion efficiency between the energetics of each single feedback event, and the release of magnetised bipolar outflows in galaxies, starting from $z=4$ and down to $z=0$.  This field is added to a negligible uniform initial seed field of $B_0 = 10^{-11}\ \mathrm{nG}$ (comoving), leading to "magnetic bubbles" correlated with halos in the simulated volume. 
 \item  \textit{``primordial+astroph''}: a model that combines the same magnetisation scheme of the "astroph" model, but it also assumes a primordial uniform magnetic field of $B_0=0.01\ \mathrm{nG}$ initialised at the beginning of the simulation.
\end{enumerate}

As an important improvement over our previous work, in these simulations we include the effect of radiative (equilibrium) cooling on baryon gas, assuming for simplicity a primordial chemical composition. This is motivated because a recent analysis of previous runs has shown that the density statistics, even in the mild density regime of the cosmic web, is more realistic when cooling is included since the start, compared to simpler non-radiative runs \citep{2022MNRAS.515..256P}.

The adopted cosmological parameters are as for Section~\ref{sec:intro}.  
The production of these new simulations was motivated in order to produce long lines-of-sight (LOS) with a finely sampled redshift evolution of gas and magnetic field quantities from $z=3$ to $z=0$, which was not available in existing simulations. 

To allow a comparison with the observed RM, we  generated 100 LOS through each simulated volume, with information of gas density and 3D magnetic field from $z=3$ to $z=0$. Each LOS is $\approx 6.1$  comoving $\rm ~Gpc$ long and was produced by replicating the simulated volume 72 times, using 21 snapshots saved at nearly equally spaced redshifts, and by randomly varying the volume-to-volume  crossing position for a total of $\approx 36800$ cells for each simulated LOS. 

We note that a second dataset of cosmological simulations, already extensively presented elsewhere \citep[e.g.][]{2017CQGra..34w4001V, 2019MNRAS.486..981G}, was used to estimate the evolution of the diameter of filaments with redshift in Sec.\ref{sec:fit_semi}, as catalogues of thousands of filaments were already available for this. The physical prescriptions in these runs were very similar to those used in our main simulations, and additional differences in the adopted numerical resolution are expected to play no role in the analysis of filament diameters derived there.

%----
\section{Evolution with redshift of  filament magnetic fields}
\label{sec:zevo}
%----

In this Section we investigate whether the RRM rms measured at different redshifts can constrain the evolution of the magnetic field in cosmic filaments. We start with considerations on the environment that generates the RRM of our sample at 144-MHz.  Then, we do  a simple, semi-analytical analysis assuming simple evolution with redshift of cosmic quantities.  We then  carry out a more accurate analysis taking a more realistic  gas density distribution from  cosmological  MHD simulations, either assuming a constant field strength or having it related to the gas density. We assume that the gas is 100 percent ionised, which is a safe assumption out to $z=5.3$ \citep{2022MNRAS.514...55B}. Finally, we estimate the RRMs predicted by the cosmological models for a comparison  with the observational results. 
\subsection{Environment}
\label{sec:env}
In Paper I we found that  an IGM origin is favoured for the RRMs of our sample at low frequency, instead of local to the source. This was inferred from the behaviour of the RRM with fractional polarization ($p$) and redshift, and the evolution of $p$ with redshift. We also found that these sources reside far from galaxy clusters at a projected distance that peaks at $\approx 5\,R_{200}$ that is well beyond a cluster virial radius $R_{100}\approx 1.36\, R_{200}$ \citep[][]{2014efxu.conf..362R}. We repeated the analysis of Paper I and found that 7 percent of the sources have a projected distance from clusters closer than $R_{100}$, which means that only $\approx 0.07^{3/2} = 2$~percent of them are estimated to have a 3D-separation shorter than   $R_{100}$ (see Appendix~\ref{app:2D3D}). 
$R_{100}$ is the distance within which the mean density of the galaxy  cluster is 100$\times$ the critical density of the Universe ($\rho_c$). From simulations, we find  this corresponds to a local overdensity of $\rho/\rho_c \approx 50$ or, in terms of mean matter density $\left< \rho_M\right>$, $\rho_M/\left< \rho_M\right> \approx 160$, according to our cosmology. 

This shows that the polarized sources are not embedded in galaxy cluster environments at these frequencies. We also checked that  intervening clusters are far from the LOS of our sources, with a similar analysis to that of Paper I. For each source of our sample, we searched for the intervening galaxy cluster with the smallest projected separation from the LOS in $R_{100}$ units.
We used the galaxy cluster catalogue of \citet{2015ApJ...807..178W}, that contains 158,103 records in the redshift range of 0.05--0.75, either spectroscopic or photometric, with an error of up to 0.018. The cluster masses are as low as $2\times 10^{12}\, \rm M_\odot$ and the  sample is 95 percent complete for  masses larger than $10^{14}\,\rm M_\odot$. For  each source at redshift $z_s$, we searched for the smallest projected separation to  the  LOS  of  the clusters at redshift $z_{gc} < z_s- 0.036$ (2-sigma uncertainty). We found that 5.2 and 8.9 percent of the sources have a LOS that passes at a distance from a cluster closer than $R_{200}$ and $R_{100}$, respectively. The median minimum projected separation is 3.5~$R_{200}$, or 2.6~$R_{100}$, which is well beyond the   cluster environment. If we restrict the search to clusters of masses larger than  $10^{14}$ $\rm M_\odot$, which are expected to  give the largest effects, those fractions  drop to 2.4 and 4.9 percent for  $R_{200}$ and $R_{100}$. These results are comparable  to those of the  analysis on the closest galaxy cluster separation, and the same considerations hold.   Only sources within the galaxy cluster catalogue footprint and in its redshift range were used, providing  739 sources for this analysis. 

\citet{2022MNRAS.515..256P}  estimated  the differential RRM of close pairs of sources from the same LoTSS RM catalogue at 144-MHz we use here, either  random pairs (rp: sources apparently close but physically separated and at different redshift) or physical pairs (pp: two components of the same source, such as two lobes of a radio galaxy, that  are   at the same distance).Differential RMs have been employed  to investigate either the magnetic field in the IGM \citep[e.g.,][]{2019ApJ...878...92V} or the ICM in galaxy clusters \citep[e.g.,][]{2022ApJ...926...65X}. For the latter the pp are used and it is   best applied  at higher frequencies where the polarized sources can populate clusters (see above). The differential RRM of a random pair  has three contributions:   the IGM intervening  the two sources;    their  local environment;  and a possible contamination from the residual GRM.  Physical pair differential RRMs have two possible contributions:  the environment local to the sources and the possible  residual GRM. Those authors measured medians of differential $|\Delta\rm RRM|$ of $\left<|\Delta \rm RRM_{rp}|\right > =1.79\pm 0.09$ rad m$^{-2}$  and  $\left<|\Delta \rm RRM_{pp}|\right > =0.70\pm 0.08$ rad m$^{-2}$ for random  and physical pairs, from which we  estimate  single source  rms  of $< \rm RRM_{rp}^2 >^{1/2} = 1.88\pm 0.09$ rad m$^{-2}$ and $< \rm RRM_{pp}^2 >^{1/2} = 0.73\pm 0.08$ rad m$^{-2}$, once we have corrected  by  1.4826 to estimate rms from the median absolute deviation\footnote{Possible in case of  zero mean, as found in Paper I} and divided by $\sqrt{2}$ to get the single source rms. The former is in excellent agreement with our estimate in Paper I of $1.90\pm 0.05$ that used single source RRMs. 

To get an estimate of the sole IGM contribution we can quadratically subtract those two values, which gives  $< \rm RRM_{\rm IGM}^2 >^{1/2} = 1.73\pm 0.12$ rad m$^{-2}$. That is only 8 percent smaller than the measured term that is thus largely dominated by the IGM RRMs. This is a further indication that our measured RRMs are mostly generated by the IGM and we will assume so in the rest of the paper. To account for the local origin contribution we add  an error of 8 percent to our  RRM rms estimates.  

We used the cosmological MHD simulations described in Section~\ref{sec:simul} to estimate the fraction of the IGM RRM that is from filaments and voids. For each cosmological model, we measured the RRM at each redshift out to $z=3$ for each of the 100 LOS using density and magnetic field from the simulations and then computed the rms  $< \rm RRM_{\rm IGM,sim}^2 >^{1/2}$. We only considered cells with  density excess $\delta_M = \rho_M/ \left< \rho_M\right> < 160$, to account for the fact that most of our  sources are far  from galaxy clusters (i.e.~estimating the RRM rms of the entire IGM excluding clusters).    We measured the $< \rm RRM_{\rm voids,sim}^2 >^{1/2}$ from voids with the same procedure, except we only considered cells with a gas density excess of $\delta_g = \rho_g/\left< \rho_{g} \right > < 1$, which is a conservative separation threshold between filaments and voids \citep{2014MNRAS.441.2923C, 2015A&A...580A.119V}. We then computed the median of the ratio  $< \rm RRM_{\rm voids,sim}^2 >^{1/2} /  < RRM_{\rm IGM,sim}^2 >^{1/2} $ for all models and we find that it is smaller than 0.013 (it ranges $1\times 10^{-3}$ to 0.013 depending on the model), for a fractional contribution of the voids to  $<\rm  RRM_{\rm IGM,sim}^2 >^{1/2} $ of less than $1\times 10^{-4}$. 
From this, we can  conclude that  voids provide  a negligible contribution to  our sample RRMs that  therefore mostly have an origin from cosmic filaments.   

\subsection{Semi-analytical analysis}
\label{sec:fit_semi}

The RRM of a source at redshift $z$ is
\begin{equation}
    {\rm RRM} = 0.812\int_z^0 \frac{n_e(z')B_\parallel (z')}{(1+z')^2} \,\frac{dl}{dz'} \, dz'
    \label{eq:rm}
\end{equation}
where the integration is performed from the source to the observer along the path length $l$ (pc), $n_e$ is the electron number density (cm$^{-3}$), and $B_\parallel$ is the magnetic field along the line of sight ($\mu$G), all referred to physical quantities.

In cosmic filaments the electron  density is $n_{e,f} = K_f\, n_e$ where $n_e$ is the average electron  density of the Universe and  $K_f$ is the filament overdensity, that is $K_{f,0}=10$ at $z=0$  \citep{2014MNRAS.441.2923C, 2015A&A...580A.119V} and evolves  as \citep[][we derived this dependency from their Figure 25]{2014MNRAS.441.2923C}
\begin{equation}
     K_f \approx K_{f,0} \, (1+z)^{-0.75}. 
     \label{eq:kf_vs_z}
\end{equation}
Hence, the electron number density in a cosmic filament varies with redshift as
\begin{equation}
    n_{e,f} = K_f\,\,n_{e,0}\,\left( 1+z \right)^3
\end{equation}
where $n_{e,0}$ is the mean  comoving (at $z=0$) electron number density of the Universe, and the RRM of a source at redshift $z$  by cosmic filaments intercepted by the source radiation is: 
\begin{equation}
    {\rm RRM_f} = 0.812\,\, K_{f,0}\,\, n_{e,0} \int_z^0 B_\parallel \,(1+z')^{0.25} \,\frac{dl}{dz'} \, dz'
    \label{eq:rmf}
\end{equation}

The medium  can be assumed to be distributed in $N_f(z)$ filaments intercepted by the LOS out to redshift $z$, and Equation~(\ref{eq:rmf}) can be written as 
\begin{equation}
    {\rm RRM_f} = 0.812 \,\,K_{f,0}\,\, n_{e,0} \sum_i^{N_f(z)} B_{\parallel,f,i}\,\,(1+z_i)^{0.25}\, \,l_f
    \label{eq:rmfsum}
\end{equation}
where $B_{\parallel,f,i}$ is $B_\parallel$ of a filament at redshit $z_i$ and $l_f = (\pi/2)\, D$ is the typical path of the LOS through a filament, considering the typical width of a filament ($D$) corrected for the average inclination to the LOS of the filament  (see Appendix B of Paper I). The typical width of a filament at $z=0$ is $D_0 \approx 6$\,Mpc \citep{2010MNRAS.408.2163A, 2014MNRAS.441.2923C,2020A&A...641A.173G}.
To estimate the evolution of $D$ with redshift, we used the statistics of filaments already extracted in a suite of simulations produced elsewhere \citep{2019MNRAS.486..981G}, with the same numerical method and (nearly) physical prescriptions of the new simulations introduced in Sec.\ref{sec:simul}. 
We  detected filaments at redshifts out to $z=3$ using an excess density threshold criterion of $\delta = 10$ at $z=0$ and  decreasing with $z$ following the growth-rate of cosmic structures \citep[see equation B5 of ][]{2011ApJ...740..102K} down to $\delta = 2.48$ at $z=3$. The fit to the mean filament radius, weighted for the filament density, provides a dependence $D\,({\rm comoving}) \propto(1+z)^{-0.4}$ and  in physical coordinates we can assume
\begin{equation}
    l_f = l_{f,0}\,\,(1+z)^{-1.4}
\end{equation}
where $l_{f,0} =  \pi/2\,\,D_0$. 

If we express $B_{\parallel,f,i} = B_{f,i}\cos \theta$, where $\theta$ is the inclination of the filament field to the LOS that is uniformly distributed over 4$\pi$-sr, and $B_f,i$ is the magnetic field strength of  filaments at redshift $z_i$, the RRM rms over all LOS can be written as: 
\begin{equation}
    \left<{\rm RRM_f}^2\right>^{1/2}= 0.812 \,\,K_{f,0}\, n_{e,0}\,l_{f,0} \,\,\sqrt{ \sum_i^{N_f(z)}\frac{ \left(B_{f,i}\,(1+z_i)^{-1.15} \right)^2}{3}}
    \label{eq:rm_rms}
\end{equation}

\begin{figure}
   \centering
    \includegraphics[width=\columnwidth]{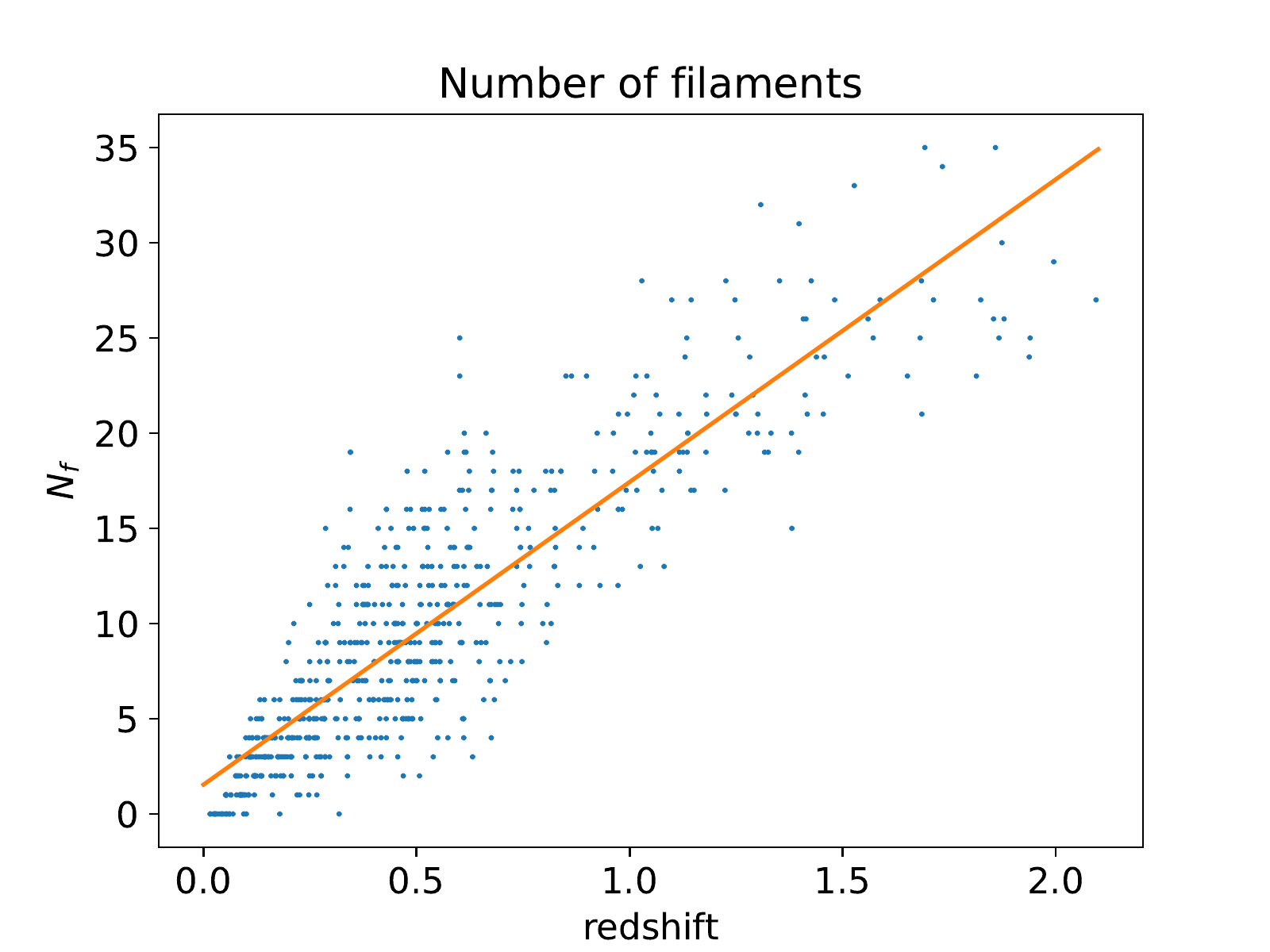}
   \caption{ Number of filaments intercepted by each of our sources in the footprint of the filaments catalogues (dots). The best-fit is also reported (solid line).}
              \label{fig:nfil}%
    \end{figure}

We can assume that  $B_f$ follows a simple power law  \citep[see][]{2022MNRAS.515..256P}
\begin{equation}
    B_f = B_{f,0} \,\,(1+z)^\alpha ,
    \label{eq:Bpowerlaw}
\end{equation}
 and, after defining 
\begin{equation}
    A_{f,0} =  0.812 \,\,\frac{K_{f,0}\, n_{e,0} \,l_{f,0}}{\sqrt{3}},
\end{equation} 
RRM$_f$ rms becomes
\begin{equation}
    \left<{\rm RRM_f}^2\right>^{1/2} = A_{f,0}\, B_{f,0}\,\, \sqrt{ \sum_i^{N_f(z)} (1+z_i)^{2\alpha-2.3} }
\end{equation}
and hence
\begin{equation}
      \left<{\rm RRM_f}^2\right>^{1/2}  = A_{f,0}\, B_{f,0}\,\, \sqrt{ \sum_i^{N_f(z)} (1+z_i)^{2\alpha-2.3}  \frac{\Delta N_f}{\Delta z}\Delta z}
\end{equation}
that can be turned into an integral
\begin{equation}
      \left<{\rm RRM_f}^2\right>^{1/2}  = A_{f,0}\, B_{f,0}\,\, \sqrt{ \int_0^{z} (1+z')^{2\alpha-2.3}  \frac{d N_f}{dz'}\,\, dz'}
      \label{eq:rm_rms_int}
\end{equation}
The number of filaments $N_f$ is, to a good approximation, linear with $z$ (see Paper I). We followed the same analysis of Paper I to estimate the number of filaments intercepted by the LOS of the sources of  our RM catalogue. We used the filament catalogues by \citet{2016MNRAS.461.3896C} and \citet{2022A&A...659A.166C} and found the number of filaments intercepted by each of the  RM catalogue sources that are in their footprint, and fit the distribution of $N_f$ so obtained (Figure~\ref{fig:nfil}). We assumed a filament width of 6\,Mpc at $z=0$ \citep{2010MNRAS.408.2163A, 2014MNRAS.441.2923C, 2020A&A...641A.173G}, evolving with redshift as discussed above.    Differing from Paper I, we   did the analysis out to the max distance of the filament catalogues ($z=2.2$), considered a width changing with redshift,  and executed a linear fit, which gives 
\begin{eqnarray}
  N_f &=& N_0 + N_1\, z\\
  N_0 &=& 1.5 \nonumber\\
  N_1 &=& 15.9 \nonumber
\end{eqnarray}
We do not have filament data  beyond $z=2.2$ and  we extrapolate this relation out to $z=3$. 

%%% Figure %%%%
%%%%%

%%%% Figure %%%%
   \begin{figure*}
   \centering
    \includegraphics[width=\textwidth]{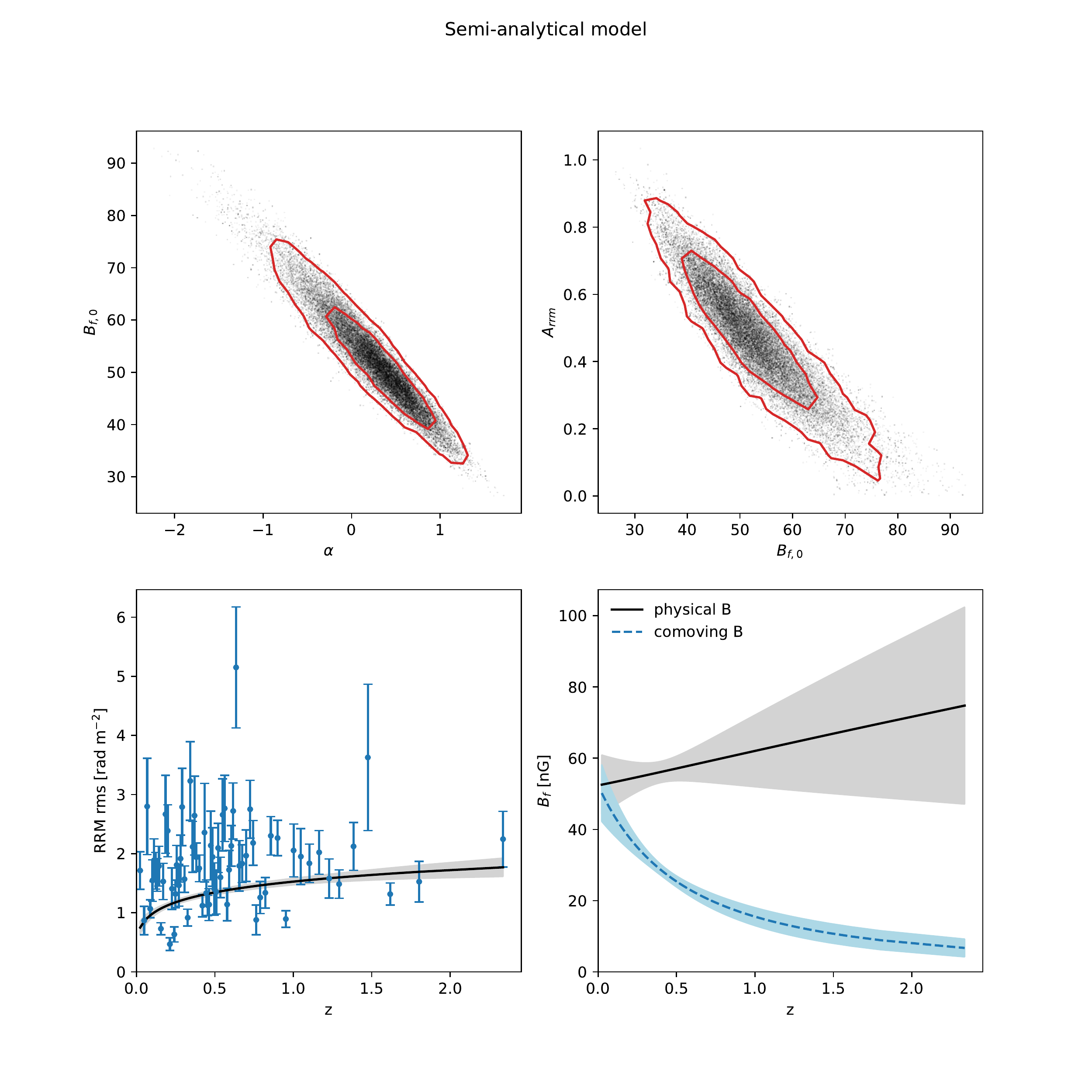}
   \caption{Best-fit results of the semi-analytical model of equations~(\ref{eq:rm_rms_a_rrmf}) and (\ref{eq:rm_rms_int3}) to the RRM rms computed with redshift bins of 15 sources per bin. {\em Top-left and Top-right}: 2D distributions (dots), and 1-sigma and 2-sigma confidence level  contours (solid lines) of the fit parameters $\alpha$, $B_{f,0}$, and $A_{rrm}$.  {\em Bottom-left}: RRM rms measured in redshift bins (circles) and best-fit curve (solid) and its error range (grey-shaded area). {\em Bottom-right}:  Evolution with redshift $z$ of the best-fit filament physical (solid line) and comoving magnetic field amplitude (dashed). The error range is also shown (shaded areas).  }
              \label{fig:rrmfit_15}%
    \end{figure*}
%%%%%

%%%%%%
% table
	%\input{tab1-old.tex}
	%%%%%%
% table
\begin{table*}
	\centering
	\caption{Best-fit parameters of the filament magnetic field  evolution with redshift  to the RRM rms measured at 144-MHz for the semi-analytical case and that with density taken from simulations. Columns are the case studied and the fit parameters:  the slope $\alpha$ of the filament magnetic field strength behaviour vs redshift; the strength $B_{f,0}$ of the filament magnetic field at $z=0$;  the constant term $A_{rrm}$;  the slope $\gamma$ of the comoving magnetic field.  Semi-analytical cases  are with  different  redshift-bin sizes. Cases with density taken from simulations are all fit to RRMs in 15-source redshift bins and differ for the magnetogenesis scenario and the overdensity  threshold of the cells used to estimate the RRMs  (matter overdensity  $\delta_M < 160$ and  $\delta_M < 100$).} 
	\label{tab:rrmf_fit}
	\begin{tabular}{ccccc}
	  \hline 
        case & $\alpha $ &  $B_{f,0}$ &  $A_{rrm}$  &   $\gamma$  \\ 
           &   & [nG]  & [rad m$^{-2}$]  &   \\ 
       \hline
       Semi-analytical model,  $l_f \propto (1+z)^{-1.4}$ \\
       \hline
       15-sources per $z$-bin &  $0.2 \pm 0.5$  &  $52 \pm 9$  &  $0.48 \pm 0.16$ &  $-1.8 \pm 0.5$ \\
       60-sources per $z$-bin &  $-0.8 \pm 2.0$  &  $80 \pm 40$ &  $1.0 \pm 0.5$ &  $-2.8 \pm 2.0$ \\
      \hline        
      density from simulations, $B_f$ constant with density,\\
       15-source $z$-bins, $\delta_M < 160$  \\
       \hline
       density model\\
       primordial uniform & $0.1 \pm 0.5$ & $47 \pm 8$ & $0.62 \pm 0.14$ & $-1.9 \pm 0.5$ \\
       dynamo & $-0.1 \pm 0.5$ & $56 \pm 10$ & $0.49 \pm 0.15$ & $-2.1 \pm 0.5$ \\
       astroph & $-0.2 \pm 0.5$ & $58 \pm 11$ & $0.57 \pm 0.16$ & $-2.2 \pm 0.5$ \\
       primordial+astroph & $0.0 \pm 0.5$ & $55 \pm 10$ & $0.64 \pm 0.13$ & $-2.0 \pm 0.5$ \\
       primordial stochastic & $-0.2 \pm 0.4$ & $57 \pm 9$ & $0.53 \pm 0.14$ & $-2.2 \pm 0.4$ \\
       \hline % mean of 55;  47-58;  39
         density from simulations, $B_f$ constant with density,\\
       15-source $z$-bins, $\delta_M < 100$  \\
       \hline
       density model &  \\
       primordial uniform & $0.0 \pm 0.5$ & $61 \pm 11$ & $0.60 \pm 0.14$ & $-2.0 \pm 0.5$ \\
       dynamo & $0.0 \pm 0.4$ & $65 \pm 10$ & $0.57 \pm 0.14$ & $-2.0 \pm 0.4$ \\
       astroph & $-0.2 \pm 0.5$ & $70 \pm 14$ & $0.56 \pm 0.16$ & $-2.2 \pm 0.5$ \\
       primordial+astroph & $-0.1 \pm 0.4$ & $69 \pm 11$ & $0.59 \pm 0.13$ & $-2.1 \pm 0.4$ \\
       primordial stochastic & $-0.1 \pm 0.6$ & $68 \pm 14$ & $0.55 \pm 0.16$ & $-2.1 \pm 0.6$ \\
       \hline % mean of 67;  61-70;  84
	\end{tabular}
\end{table*}
%%%%%%

%%%%%%
% table
\begin{table*}
	\centering
	\caption{Best-fit parameters of the filament magnetic field  evolution with redshift  to the RRM rms measured at 144-MHz for the case with density taken from simulations and magnetic field dependent on the gas density as $B_f \propto \rho_g^{2/3}$. Columns are the case studied and the fit parameters:  the slope $\alpha$ of the filament magnetic field  behaviour vs redshift; the strength $B_{f,0}^{10}$ of the filament magnetic field at $z=0$ and gas overdensity $\delta_g =10$;  the constant term $A_{rrm}$;  the slope $\gamma$ of the comoving magnetic field.  All cases are fit to RRM rms computed in 15-source redshift bins and differ for the magnetogenesis scenario. Two sets of cases are reported, differing for the overdensity threshold of the cells used to estimate  the  RRMs from simulations ($\delta_M < 160$ and  $\delta_M < 100$).} 
	\label{tab:rrmf_fit_Bdens}
	\begin{tabular}{ccccc}
	  \hline 
        case & $\alpha $ &  $B_{f,0}^{10}$ &  $A_{rrm}$  &   $\gamma$  \\ 
           &   & [nG]  & [rad m$^{-2}$]  &   \\ 
       \hline
       $B_f \propto \rho_g^{2/3}$, $\delta_M < 160$ \\
       \hline
       density model &  \\
       primordial uniform & $0.3 \pm 0.5$ & $10.0 \pm 2.0$ & $0.70 \pm 0.14$ & $-1.7 \pm 0.5$ \\
       dynamo & $0.1 \pm 0.4$ & $11.8 \pm 1.9$ & $0.55 \pm 0.15$ & $-1.9 \pm 0.4$ \\
       astroph & $-0.1 \pm 0.5$ & $12.8 \pm 2.4$ & $0.61 \pm 0.15$ & $-2.1 \pm 0.5$ \\
       primordial+astroph & $-0.1 \pm 0.4$ & $12.8 \pm 2.1$ & $0.65 \pm 0.13$ & $-2.1 \pm 0.4$ \\
       primordial stochastic & $-0.1 \pm 0.5$ & $14.2 \pm 2.6$ & $0.47 \pm 0.17$ & $-2.1 \pm 0.5$ \\
       \hline % mean of 12.3;  10.0--14.2;  8
       $B_f \propto \rho_g^{2/3}$,  $\delta_M < 100$ \\
       \hline
       density model &  \\
       primordial uniform & $0.0 \pm 0.6$ & $18.0 \pm 3.9$ & $0.63 \pm 0.16$ & $-2.0 \pm 0.6$ \\
       dynamo & $0.1 \pm 0.4$ & $18.6 \pm 3.1$ & $0.60 \pm 0.14$ & $-1.9 \pm 0.4$ \\
       astroph & $-0.2 \pm 0.5$ & $20.9 \pm 3.6$ & $0.57 \pm 0.15$ & $-2.2 \pm 0.5$ \\     primordial+astroph & $0.3 \pm 0.5$ & $17.2 \pm 3.1$ & $0.61 \pm 0.14$ & $-1.7 \pm 0.5$ \\
       primordial stochastic & $-0.2 \pm 0.5$ & $22.4 \pm 3.9$ & $0.48 \pm 0.15$ & $-2.2 \pm 0.5$ \\
       \hline % mean of 19;  17--22; 26
	\end{tabular}
\end{table*}
%%%%%%
% end table
%%%%%%

Equation~(\ref{eq:rm_rms_int}) thus becomes
\begin{equation}
      \left<{\rm RRM}^2\right>^{1/2}  = A_{f,0}\, N_1^{1/2}\, B_{f,0}\,\, \sqrt{ \int_0^{z} (1+z')^{2\alpha-2.3} \,\, dz'}
      \label{eq:rm_rms_int2}
\end{equation}
with solution
\begin{equation}
      \left<{\rm RRM_f}^2\right>^{1/2} =
      \begin{cases}
           A_{f,0}\, N_1^{1/2}\, B_{f,0}\,\sqrt{ \frac{(1+z)^{2\alpha-1.3}-1}{2\alpha-1.3}} &{\rm for}\,\,\alpha\neq 0.65 \\ 
           \\
           A_{f,0}\, N_1^{1/2}\, B_{f,0}\,\sqrt{ \ln (1+z)}& {\rm for}\,\,\alpha = 0.65  \label{eq:rm_rms_int3}
      \end{cases}
\end{equation}
As discussed in Section~\ref{sec:env}, a cosmic filament origin can be assumed for our RRM sample measured at  144-MHz with LOFAR.  Hence, we fit the measured RRM rms to the function 
\begin{equation}
     \left<{\rm RRM}^2\right>^{1/2} = \frac{A_{rrm}}{(1+z)^2} +  \left<{\rm RRM_f}^2\right>^{1/2}
     \label{eq:rm_rms_a_rrmf}
\end{equation}
where, besides the cosmic filament term RRM$_f$, we allow a constant RRM term, corrected for redshift, to account for a possible additional contribution different from  filaments. This is motivated because the model  RRM$_f$ converges to zero at $z=0$ while our measured RRMs  do not.   
We use  a Bayesian fit\footnote{{\small EMCEE} package \citep{2013PASP..125..306F}: https://pypi.org/project/emcee/}, with priors of $B_{f,0} < 250$~nG \citep[][]{2021A&A...652A..80L}, $B_{f,0} \geqq 0$, and $A_{rrm} \geqq 0$.  The results are shown in Figure~\ref{fig:rrmfit_15} where we used the RRM rms computed in bins of 15 sources each. The 2D confidence level contours of the parameters, the best-fit model, and the resulting evolution of $B_f$ with redshift are shown. The best-fit results (Table~\ref{tab:rrmf_fit}) give a filament magnetic field with a slope  $\alpha=0.2 \pm 0.5$ that is consistent with no evolution with redshift, as also shown in  Figure~\ref{fig:rrmfit_15}, bottom-right panel, and with an amplitude at  $z=0$  of $B_{f,0} = 52 \pm 9$~nG. The latter can be also written as: 
\begin{equation}
    B_{f,0} = (52 \pm 9) \,\left( \frac{6\,\,{\rm Mpc}}{D_0} \right) \,\,{\rm~nG}
\end{equation}
that shows  the  dependence of $B_{f,0}$ on the filament width (these two parameters are inversely proportional, see Equation~(\ref{eq:rm_rms_int3})). 

Assuming the magnetic field is frozen to the plasma, the magnetic field goes as $n_e^{2/3}$ and thus $B_f = B_{f,c}\, (1+z)^2$, where  $B_{f,c}$ is the comoving magnetic field in filaments that, according to our model for $B_f$, varies with $z$ as

\begin{equation} 
    B_{f,c}(z) = B_{f,0} \, (1+z)^\gamma\,\,\,\rm   with \,\,\gamma = \alpha - 2. 
\end{equation}

From the results of our fit, hence, we get $\gamma =-1.8 \pm 0.5$, which gives  a comoving magnetic field in filaments that significantly evolves, decreasing with redshift.  
The behaviour of the comoving field is shown in Figure~\ref{fig:rrmfit_15}, bottom-right panel. 

We also run the fit  to the RRM rms computed with a set of  larger redshift bins (60 sources each). The results, shown in Table~\ref{tab:rrmf_fit}, are consistent with those obtained with smaller bins, albeit  with larger errors. 

\subsection{Analysis with densities from cosmological simulations }
\label{sec:fit_sim}
The semi-analytical approach is powerful and gives an insight into the terms at play, but it has limitations. Those most obvious are the gas density assumed to follow that of the dark matter and the overdensity assumed to be constant within a filament and for all filaments. To overcome this, and to obtain more precise estimates of the evolution of the field,  we make direct use of the gas density from  cosmological simulations (see Section~\ref{sec:simul}).  The goal is still to find the evolution with redshift of the mean filament magnetic field strength assuming the power law behaviour  of Equation~(\ref{eq:Bpowerlaw}). 

For each of the cosmological models considered, we extracted 100 LOS out to $z=3$. For each  LOS we calculated the RRM$_f$ at each $z$ using Equation~(\ref{eq:rm}) and the gas density from the simulation. We considered only cells  with a matter  density excess $\delta_M < 160$ to account for our sources residing far from galaxy clusters. We excluded cells with gas excess density $\delta_g <1$ because they give a negligible contribution to the total RRM, as shown in Section~\ref{sec:env}. 

The magnetic field was estimated assuming a value of $B_{f,0}$ and $\alpha$ (Equation~\ref{eq:kf_vs_z}). Each time the LOS entered a region with $\delta >1$, the direction of the magnetic field to the LOS was changed, randomly picked within 4$\pi$ sr. That ensured a magnetic field with constant orientation within each filament and randomly changing filament to filament. We did 120 realisations of these magnetic field configurations,  for a total of 12,000 realisations (100 LOS $\times$ 120 magnetic field configurations).

From these 12,000 realisations of RRM$_f$ we computed the RRM$_f$ rms that is that expected given the assumed values of $B_{f,0}$ and $\alpha$. Since we are not interested to small scale variations, we smoothed the RRM$_f$ rms with a top hat filter of width $dz=0.1$, which further reduces  the statistical variations of the individual RRM$_f$ realisations. We computed this all for different values of $\alpha$ (at the same value of $B_{f,0}$), spanning the range [-5, 5] with steps of 0.5, which covers the range of interest. We also tested a step of 0.25 with similar results. A linear interpolation between the two nearest values gives the RRM$_f$ rms estimate at any other $\alpha$ value.  The RRM$_f$ has a simple linear dependence on $B_{f,0}$ that, combined with the interpolation over $\alpha$, gives us  the functional dependence on these two parameters required by a Bayesian fit. 

The results of a Bayesian fit to Equation~(\ref{eq:rm_rms_a_rrmf}) are reported in Table~\ref{tab:rrmf_fit}, with the RRM$_f$ estimated as discussed above, and the RRMs measured in 15-source bins, for all of the cosmological models we considered. We assumed the same priors as for the semi-analytical analysis. 

The best-fit values of $\alpha$ are in the range [-0.2, 0.1] and are all consistent within the errors ($\sigma_\alpha =0.4$--0.5) and consistent with no evolution with redshift.  They are also consistent with the  value derived by the semi-analytical analysis.  The comoving magnetic field slope $\gamma$ is in the range [-2.2, -1.9], which confirms a decrement with increasing redshift. 

The mean amplitude of the field in a filament at $z=0$ is $B_{f,0} \approx 55$~nG with variations depending on the models (it ranges from 47--58 nG), but within the uncertainty that is better than 5-sigma. It is consistent with the result of the semi-analytical model.  

We regard the results obtained here as more accurate than those of the semi-analytical model, because of the better description of the gas density. However, the proximity of the results tells us that the semi-analytical model is a good approximation  and suggests that is an effective (and computationally cheaper) approach to apply to large datasets.

Currently, we cannot exactly set the limit on the overdensity. From the distribution of the source separation to the nearest cluster found in Paper I (see also Section~\ref{sec:env}) it is possible that they reside at  overdensities lower than $\delta_M=160$. Therefore, we repeated the analysis, setting the limit to $\delta_M  < 100$, which is the threshold separating  filaments and halos. This is sort of an extreme case because it assumes that all  sources are in filaments.

The   best-fit results are reported in Table~\ref{tab:rrmf_fit}. The slopes are similar to the $\delta_M=160$ case,  with small changes that are  well within the uncertainties. However, the field strength  is larger.  It ranges from 61--70~nG, with a mean value of 67~nG. In summmary, changing the overdensity limit does not affect the best-fit slope, while it changes the field amplitude, increasing it by $\approx$20 percent when changing from $\delta_M =160$ to 100. Considering all of the models, overdensity limits, and 1-sigma  uncertainties, the  magnetic field strength of a filament at $z=0$ is in the range 39--84~nG.

\subsection{Analysis with density from cosmological simulations and magnetic field frozen to matter.}
The approach used above, in Section~\ref{sec:fit_sim}, assumes a $B_f$ which is constant with density.  This results in an average value where filaments at higher density contribute more, because the  RRM of a filament  depends on $\rho_g^{5/3}$. 

Therefore, we have repeated the same analysis assuming that the magnetic field strength is
\begin{equation}
    B_{f} = B_{f,0}^{10} \left(\frac{\delta_g}{10}\right)^{2/3} \,\, (1+z)^\alpha
    \label{eq:Bf_density}
\end{equation}
where $ B_{f,0}^{10}$ is the average magnetic field strength of a filament of gas overdensity $\delta_g=10$ at $z=0$ and  $B_{f,0}^{10}\,\, (1+z)^\alpha$ is that at redshift $z$. The dependence on $\delta_g$ assumes that the magnetic field is frozen to the ionised medium\footnote{We assume there is  no further amplification, such as  by turbulent gas motions, which   numerical simulations suggest may occur \citep[e.g.,][]{2019MNRAS.486..981G}.} 
and  also seems to hold in more evolved environments like those  of  galaxy clusters \citep[e.g.,][submitted]{2022arXiv220604697R}. 
Hence, this approach estimates the field strength at the typical density of a filament ($\delta_g=10$, see  \citealt{2014MNRAS.441.2923C, 2015A&A...580A.119V}) and, following Equation~(\ref{eq:Bf_density}), at any density. 

The results of the  Bayesian fit are given in Table~\ref{tab:rrmf_fit_Bdens} and Figure~\ref{fig:rrmfit_sim} of  Appendix~\ref{app:bestfit} (for the primordial stochastic model only, the other models show similar results).  
The best-fit values of $\alpha$ are in the range [-0.1, 0.3], are all consistent within the errors ($\sigma_\alpha =0.4$--0.5), and are consistent with no evolution with redshift.  The comoving magnetic field slope $\gamma$ is in the range $[-2.1, -1.7]$, which gives a decrement with redshift also in this case. 
The mean amplitude of the field in a filament at overdensity $\delta_g = 10$ and $z=0$ is $B_{f,0}^{10} \approx  12.3$~nG (values are in the range 10.0--14.2~nG and partly depend on  the best-fit slope -- there is some degeneracy between slope and field strength, as shown by Figure~\ref{fig:rrmfit_sim}), with  uncertainties better than 5-sigma.

The typical density of a filament evolves as for Equation~(\ref{eq:kf_vs_z}). Combined with our results and the assumed relation of $B_f$ with  density, we get that the typical magnetic field of a filament evolves with redshift as $B_{f,t} = B_{f,0}^{10} \,\,(1+z)^\eta$, with $\eta = \alpha - 0.5$ ($\eta_c = \gamma -0.5$ for the comoving field), that hence  runs in the range [-0.6, -0.2] ([-2.6, -2.2] for the comoving field).

As for the case of $B_f$ constant with density, we repeated the analysis for an overdensity limit of $\delta_M = 100$. Results are reported in Table~\ref{tab:rrmf_fit_Bdens}.  The slopes are similar to the $\delta_M=160$ case. The magnetic field strength $B_{f,0}^{10}$ ranges from 17--22~nG with a mean value of $\approx$19~nG, which is $\approx$50 percent larger than the previous case. Considering all of the models, overdensity limits, and 1-sigma  uncertainties, the  magnetic field strength of a filament at $z=0$ and gas overdensity $\delta_g =10$ is in the range 8--26~nG.

\subsection{Predictions from simulations}
%

%%% Figure %%%%
   \begin{figure*}
   \centering
    \includegraphics[width=\columnwidth]{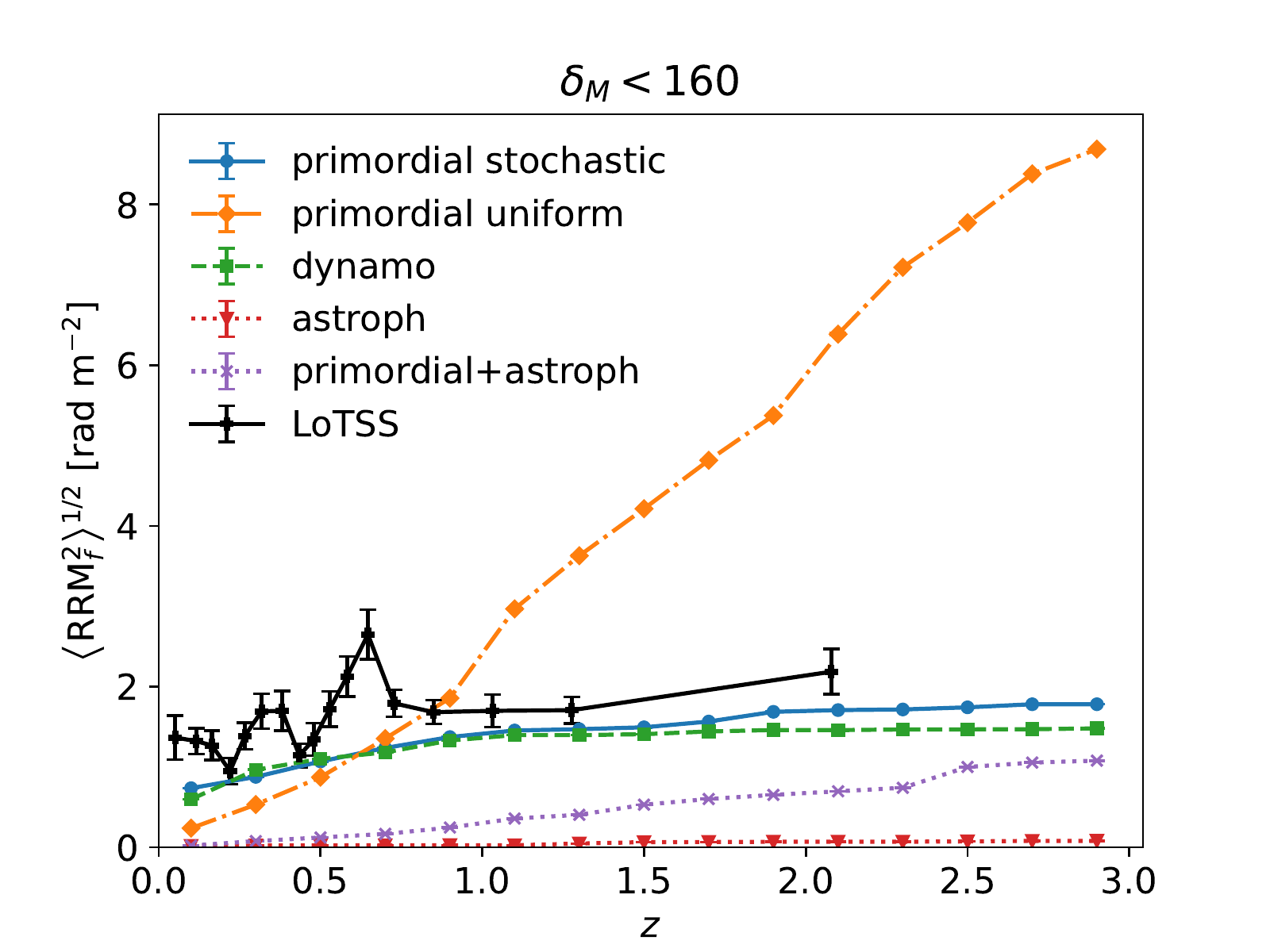}
    \includegraphics[width=\columnwidth]{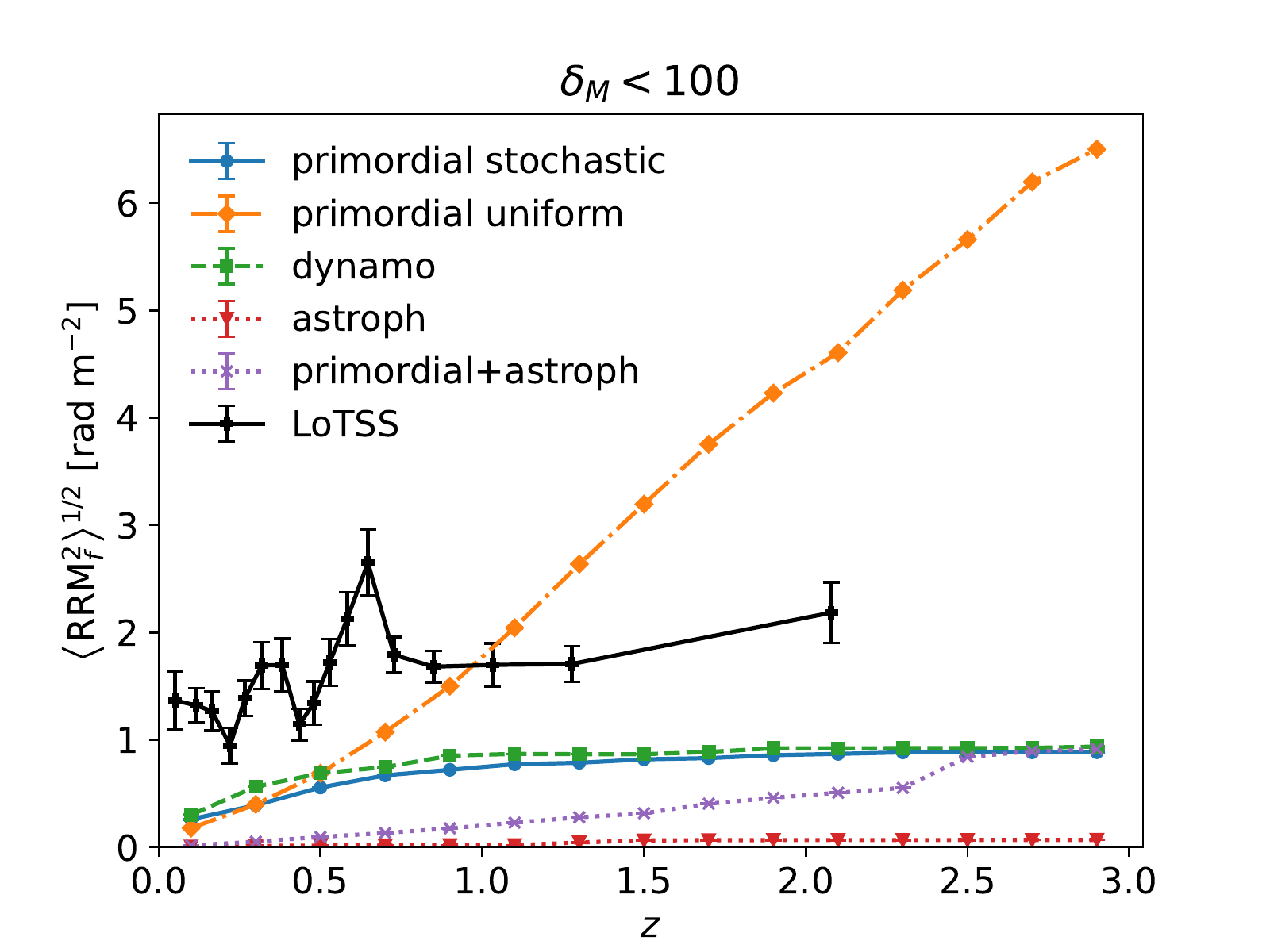}
   \caption{RRM$_f$ rms in redshift bins computed using density and magnetic field from simulations, for the magnetogenesis models we considered. Two cases differing by the  overdensity limit are shown: $\delta_M < 160$ (left) and $\delta_M < 100$ (right). The  RRM rms  from the LoTSS RM catalogue measured in 60-source redshift bins is also shown, with the term  $A_{rrm}(1+z)^{-2}$ subtracted off.}
              \label{fig:rrmf_sim}%
    \end{figure*}
%%%%%

%%% Figure %%%%
   \begin{figure}
   \centering
    \includegraphics[width=\columnwidth]{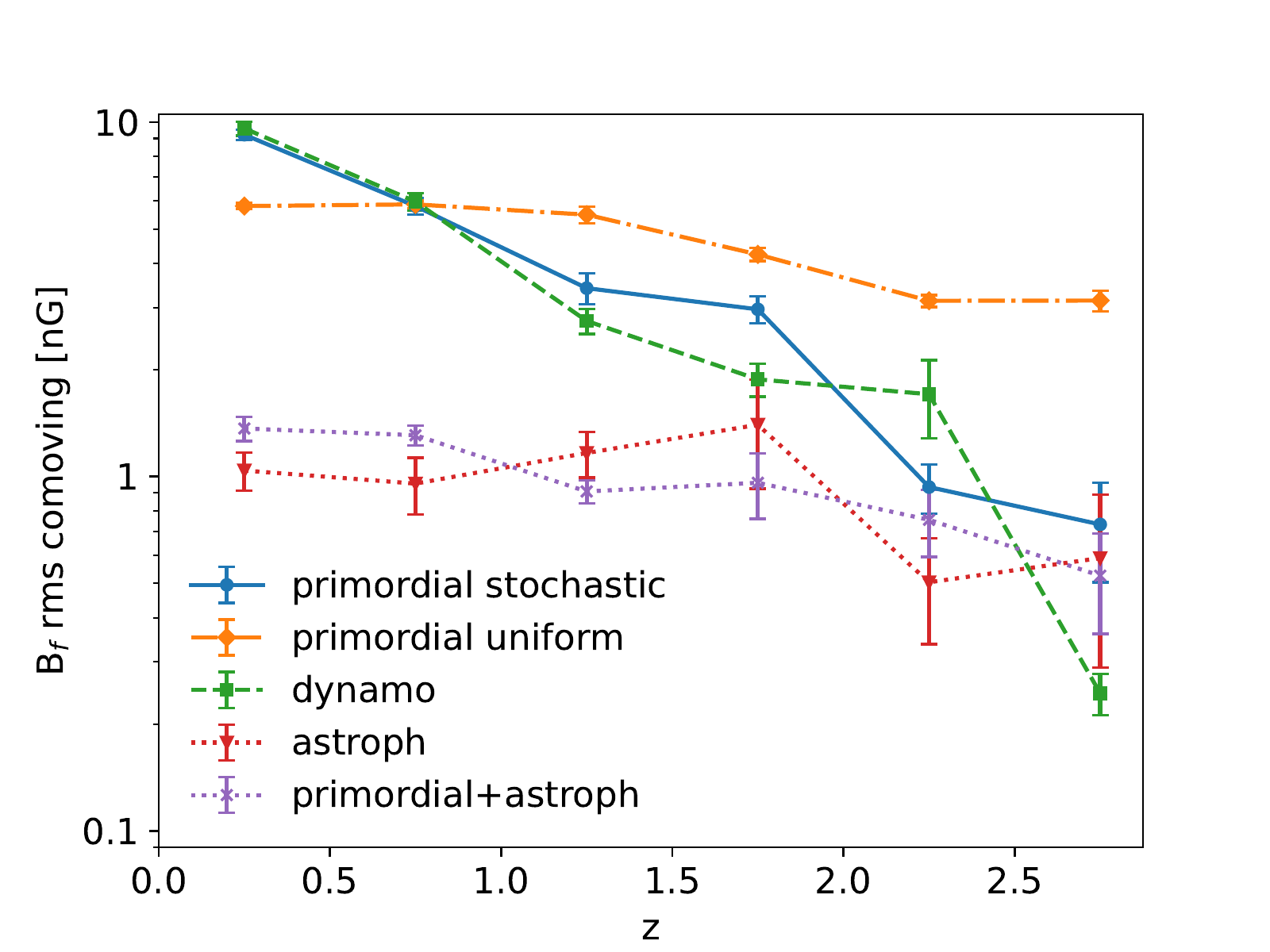}
   \caption{Comoving magnetic field strength at overdensity $\delta_g =10$ estimated from the simulations of the  magnetogenesis models considered in this work.  }
              \label{fig:Bf_sim}%
    \end{figure}
%%%%%

%%%%%%

%%%%%%
% table
\begin{table*}
	\centering
	\caption{Results of the power-law best-fit to the comoving  magnetic field at $\delta_g =10$ for all of the cosmological models we considered. The power law is  $B_{f,c}^{10} = B_{f,0}^{10}\,\,(1+z)^\gamma$ and the columns are:  model;  best-fit of the  slope $\gamma$ and  amplitude $B_{f,0}^{10}$ (field strength at $z=0$ and $\delta_g=10$);   difference between these best-fit values and those from the fit to the observations in Table~\ref{tab:rrmf_fit_Bdens} in the case of an overdensity threshold of $\delta_M < 160$ ($\Delta \gamma$ and  $\Delta B_{f,0}^{10}$). } 
	\label{tab:Bf10_sim}
	\begin{tabular}{ccccc}
	  \hline 
        model & $\gamma $ &  $B_{f,0}^{10}$   & $\Delta \gamma $ &  $\Delta B_{f,0}^{10}$    \\ 
           &   & [nG]  &  &  [nG]  \\ 
       \hline
       primordial uniform  &  $-0.6\pm 0.2$ & $7.7 \pm 1.2$  &  $1.1 \pm 0.5$  &  $-2\pm 2$ \\
       dynamo              &  $-2.8 \pm 0.7$  &  $25 \pm 15$  &   $-0.9\pm 0.8$  &  $13\pm 15$\\
       astroph             &  $-0.5 \pm 0.4$  &  $1.4\pm 0.5$ &  $1.6 \pm 0.6$ & $-11 \pm 2$\\ 
       primordial+astroph  &  $-0.8\pm 0.2$  & $1.8 \pm 0.3$ &  $1.3 \pm 0.4$ & $-11\pm 2$\\
       primordial stochastic & $-2.3 \pm 0.4$ & $20 \pm 7$  & $-0.2\pm 0.6$ & $6\pm 7$\\
       \hline
	\end{tabular}
\end{table*}
%%%%%%

It is difficult to discriminate between the different magnetogenesis scenarios based solely on the results of the previous subsections, because those scenarios do not differ  much in  gas density and  give   similar outcomes 
(which, however, gives us a nearly model-independent estimate of the evolution of $B_f$). 
However, we can  use their directly simulated prediction of the RRM and $B_f$ evolution with redshift. 

The RRM$_f$ rms of the IGM were computed for all of the cosmological scenarios following the procedure of Section~\ref{sec:env}, using the density and magnetic field values from  the simulations  and  considering only the LOS cells with an overdensity under a given limit ($\delta_M < 160$ or $\delta_M < 100$). The results are shown in Figure~\ref{fig:rrmf_sim} for both overdensity limits. The  RRM rms measured from our sample  in 60-source bins, that have lower errors for an easier comparison,  is also displayed after subtracting off the term $A_{rrm}(1+z)^{-2}$  to  show the component that the fit attributes to the IGM only. We set $A_{rrm} =0.6$ rad m$^{-2}$ which is an intermediate value  of the best-fit results. 

The observed RRMs are  best matched in both shape and amplitude (for the $\delta_M=160$ case) by  the dynamo and  primordial stochastic models, whose RRM rms flattens at high redshift. The others look disfavoured. In particular, the astrophysical model predicts RRMs that are too small, while the RRM of the  mixed primordial uniform+astrophysical  and primordial uniform models increases nearly linearly with redshift out to $z=3$. 

The comoving magnetic field strength at $\delta_g =10$  is estimated as the rms of   the magnetic field from  cells  in the 100 LOS with $\delta_g$ in the range 2--50, and is  shown in Figure~\ref{fig:Bf_sim}. We also tried narrower ranges, but the statistics were too poor and  the results were unstable. Table~\ref{tab:Bf10_sim} reports the results of a best fit to a power law $B_{f,c}^{10} = B_{f,0}^{10}\,\,(1+z)^\gamma$. 
To facilitate the comparison we report the differences with the results of the best fits to the observations for our fiducial overdensity limit of $\delta_M =160$.  Comparisons with the slopes in the case of a limit of  $\delta_M =100$ are similar.  The dynamo and primordial stochastic models only posses a comoving slope consistent with the results from the observations. This is not surprising because a steep slope is required to obtain a RRM rms that flattens at high redshift. The  magnetic field strength range allowed by the observations is broad. However, the dynamo and primordial stochastic models appear to predict strengths consistent with such a range, albeit the former has a large uncertainty.

%----
\section{Discussion}
\label{sec:discussion}
%----
Our analysis indicates  that the RRM rms we used is dominated by the component originated in cosmic filaments. We find that the filament   physical magnetic field is consistent   with no evolution with redshift, regardless of the three types of analysis we have applied and the overdensity limits that we set. The comoving field decreases with redshift with slope $\gamma \approx -2.0\pm 0.5$, with small variations  depending on the magnetogenesis model. 
This is the first estimate of the evolution of the magnetic field in filaments, to our knowledge, and complements   the result by \citet{2022MNRAS.515..256P} for the average field of the  IGM, who find a slope of $\gamma_{\rm IGM} \approx -4.5$. The difference is because in filaments the evolution of their  overdensity  and   transversal size have to be added to the equation, which makes the filaments' $\gamma$ flatter by 2.15 (see Section~\ref{sec:fit_semi}) and the two results consistent. This is the direct result of a nearly flat evolution  of RRM that requires a decreasing comoving field.

The strength at $z=0$ of the filament magnetic field averaged over all filaments  is estimated at $B_{f,0} \approx 55$~nG for our fiducial value of the overdensity limit of $\delta_M =160$, with an uncertainty better than 5-sigma. Considering variations due to both overdensity limits, the fit uncertainties, and the different magnetogenesis scenarios, $B_{f,0}$ is in the range 39--84~nG.  This result is in agreement both with previous upper limits \citep{2017MNRAS.467.4914V, 2017MNRAS.468.4246B, 2019A&A...622A..16O, 2021MNRAS.503.2913A, 2021A&A...652A..80L} and  the estimate of\footnote{It  yields  $\approx 60$~nG, if equipartition is assumed, or $\approx 30 \rm ~nG$, based on  numerical simulations, albeit with a dependence on the (unknown) amount of accelerated radio emitting cosmic ray electron.} 30--60~nG obtained from the first claimed detection of the stacked synchrotron emission from filaments of the cosmic web \citep{2021MNRAS.505.4178V}.  Both our and their method   measure the field averaged over all types of filaments and are possibly dominated by the largest of them. 
Note that our result is larger than that of $\approx$30~nG derived in  Paper I. There we assumed no evolution (with an average filament density at $z=0.7$) and a filament width constant with redshift (instead of decreasing), which led to a lower field strength estimate. Adding the evolution with redshift was  thus essential to get an improved estimate.

We also estimated the magnetic field  at the typical filament overdensity of $\delta_g =10$, finding a strength of $B_{f,0}^{10}\approx 12.3$~nG at $z=0$ for our fiducial overdensity limit, and a slope with redshift similar to the previous case. If we consider all types of variations (model, overdensity limit, and fit uncertainties) the field strength is in the range 8--26~nG. 

The dynamo and  primordial stochastic models predict   RRM rms values that  flatten at high redshift and provide the best match to the observed RRM  in both shape and amplitude. The others look disfavoured. In particular,  the RRM of the  primordial uniform model increases almost linearly with redshift and  it is striking to notice how such an observed redshift evolution of RRMs clearly disfavours it. 
While a comoving uniform seed field of $\approx 0.1-0.5 {\rm ~nG}$ was shown to  be compatible with previous observations or the non-detection of synchrotron radio emission from the cosmic web \citep[e.g.][]{2017MNRAS.467.4914V,2021MNRAS.505.4178V,2021A&A...652A..80L} or more local analysis of the RRM \citep[e.g.][]{2019A&A...622A..16O}, these redshift dependent constraints on the RRM exclude such a  simplistic model of the magnetic field with high confidence. This is because even such a weak 0.1~nG magnetic field correlated on scales as long as the entire tested cosmic volume\footnote{The  initial uniform orientation of the magnetic field is preserved while evolving, on average.} of $\approx$6~Gpc  produces a systematic increase of the RRM with redshift, which is unobserved. 
This makes the RRM evolution with redshift an extremely powerful probe of cosmic magnetism on cosmic scales. 

Furthermore, the  comoving magnetic field  at $\delta_g =10$  favours the dynamo and primordial stochastic models, as they are the only models that posses both a comoving slope and strength that are consistent with the results of the best fits to the observations. 

The dynamo and  primordial stochastic models are thus favoured by our RRM measurements. The primordial uniform, mixed, and astrophysical models appear to be excluded. The former  two  predict a continuously increasing RRM rms out to $z=3$. The latter predicts RRMs that are too small and magnetic fields that at $z=0$ are one order of magnitude weaker than the results of our best fits and  a redshift  evolution that is too flat. These results are in agreement with \citet{2021MNRAS.500.5350V}, who found the stochastic model consistent with previous observational constraints,  and  \citet{2022MNRAS.515..256P}, who found the dynamo model  consistent with the evolution  of the mean IGM magnetic field.

Using previous observational constraints, \citet{2021Galax...9..109V} and  \citet{2021MNRAS.505.4178V} also  found that  the dynamo model is challenged, which, combined with  our results, favours  the primordial stochastic model only. We note that combinations of the primordial stochastic with other models are possible, but exploring  such  lines of investigation is beyond the scope of this work.

If we restrict the analysis to this most favoured primordial stochastic model, then the filament magnetic field strengths at $z=0$ are restricted to $B_{f,0} = 48$--82~nG and $B_{f,0}^{10} = 11$--26~nG, while the slopes are $\alpha =-0.15\pm0.5$ and $\gamma =-2.15\pm 0.5$. 

The  RRM rms amplitude of the primordial stochastic scenario depends linearly on the initial field strength $B_{\rm Mpc}$.  A best fit of the amplitude of the directly simulated RRM rms of Figure~\ref{fig:rrmf_sim} to the observed RRM rms   gives the value that is most consistent with our data. We found that it is $B_{\rm Mpc} = 0.051\pm 0.010 \rm ~nG$ and $ 0.097\pm 0.010 \rm ~nG$, comoving, for overdensity limits of $\delta_M =160$ and 100, respectively. We restricted the fit to the values at $z>1$ because they are least affected by the $A_{rrm} \,(1+z)^{-2}$ correction term. A range of  $B_{\rm Mpc} = 0.04$--0.11~nG, comoving,  thus best matches the RRMs of our sample, for a primordial stochastic scenario with a spectrum of slope $\alpha_s =1$, which we simulated. These results are consistent with previous upper limits of 0.12--0.13~nG  derived from CMB observations for the same scenario \citep[][]{2019JCAP...11..028P, 2022arXiv220406302P}. 

It is worth noting that the mass resolution of the simulations we ran does not allow us to reproduce the total distribution of low mass galaxies (e.g. dwarf galaxies), which can introduce an additional magnetisation baseline even in voids \citep[e.g.][]{2013MNRAS.429L..60B}. 
Simulations with  higher resolution and adaptive mesh that are better suited to resolve the formation of small galaxies in voids predict the formation of  "magnetisation bubbles". These bubbles typically have $\geq 10^{-3} \rm~nG$ fields, yet with volume filling factors and magnetisation amplitudes from dwarf galaxies that depend on the assumed input magnetic field \citep[][]{2021MNRAS.505.5038A}, which currently has an unclear contribution to the observed RRM  \citep[][]{2022arXiv220406475A}. Higher resolution simulations are thus required in future work to complete the assessment of the astrophysical scenario, although the magnetic field strength at $\delta_g=10$ in such high resolution  simulations is comparable to the $\approx 1~\rm~nG$ that we find in ours \citep[see Figure 2 of][]{2022MNRAS.515..256P} and large variations from our results are not expected.

%----
\section{Conclusions}
\label{sec:conc}
%------

We estimated the extragalactic RM contribution  (RRM) of   the RM catalogue derived from  LoTSS DR2  survey data, and, following the procedure of Paper I, measured their rms in redshift bins of sources out to $z=3$.  We used the RRM rms to investigate the evolution with redshift of the magnetic field strength in cosmic web filaments. Our main findings are: 
\begin{enumerate}
    \item The RRM component that originates local to the source   contributes only $\approx$8 percent to the total RRM. Using cosmological simulations, we  also found that  voids are expected to have a  marginal contribution to the total RRM from the IGM. The polarized radiation from our sample  at 144~MHz tends to avoid intervening galaxy clusters  along the line-of-sight. Cosmic filaments are  hence the dominant term of our observed RRMs measured at 144~MHz.
    \item Adding an error term to account for the small local origin component, we used densities from cosmological MHD simulations of five different magnetogenesis scenarios to fit a physical magnetic field in cosmic filaments of  shape $B_f = B_{f,0}\,\,(1+z)^\alpha$  to the measured RRM rms. We also  allowed  an additional constant term evolving with redshift.  In the cases where we fit a mean magnetic field,  we find the slope  is in the range $\alpha = [-0.2,0.1]$, depending on the scenario,  with an error of $\sigma_\alpha = 0.4$--0.5, which is consistent with no evolution. The comoving field has slope $\gamma = [-2.2, -1.9]$, which means that it decreases at high significance. This is as a consequence of the nearly flat behaviour of the RRM rms. The strength at $z=0$ is in the range $B_{f,0} = 39$--84~nG and is consistent with previous results based on synchrotron emission stacking.
    \item If we assume that the magnetic field depends on the gas density as $B_f \propto \rho_g^{2/3}$ (i.e. frozen to the plasma), the slopes are mostly similar to the previous case and the strength, at $z=0$ and at an overdensity of $\delta_g = 10$ that is typical of filaments, is $B_{f,0}^{10} =8$--26~nG. 
    \item  Comparing the RRM rms and $B_{f,0}^{10}$ predicted by the five simulated scenarios with those from our measurements and best fits, leads to the dynamo and primordial stochastic models being favoured, mainly  because of the flat RRM rms they predict. The primordial uniform, astrophysical, and mixed models appear to be rejected, in particular the former is disfavoured by its RRM rms that is  continuously increasing with redshift. The strong rejection of the simple primordial uniform model is a new result that is mostly due to the constraints from the evolution with redshift of the RRM rms.   Considering earlier work also, only the primordial stochastic scenario (with a spectrum of slope $\alpha_s=1$) is favoured. Its best-fit slope is $\alpha =-0.15\pm0.5$. The comoving field slope is  $\gamma =-2.15\pm 0.5$. 
    The best-fit value of the initial field is  $B_{\rm Mpc} = 0.04$--0.11~nG.
\end{enumerate}
This work has provided a first advance  of our initial analysis conduced in Paper I and has led us to estimating the behaviour with redshift of the magnetic field in cosmic web  filaments. This has  thus also  provided a more accurate estimate of the field strength.   We find  that the physical field is consistent with no evolution and the comoving field decreases with redshift with a slope $\gamma \approx -2.0\pm 0.5$ ($-2.15\pm 0.5$ for the most favoured scenario). Such a result is because of the nearly flat RRM rms behaviour with redshift, and has important implications on understanding what  process has generated magnetic fields in the Universe and how they have evolved. A primordial field with a uniform initial field is unsuitable. A primordial field with random stochastic initial conditions is favoured and we find a range of initial field strengths that best match our data.

Further advances can be pursued with future work and data.  An improvement of a factor of three of the RRM rms uncertainties, that could be reached with 9$\times$ more sources, would give errors on $\alpha$ of $\approx$0.15, significantly  improving the precision. This is within the reach of the full LoTSS survey that will have larger area (4$\times$), better resolution (6 versus 20 arcsec), and improved polarized source selection (see discussion in O'Sullivan et al.  submitted). ASKAP-POSSUM \citep{2010AAS...21547013G}, APERTIF \citep{2022A&A...663A.103A}, and SKA-LOW  \citep{2019arXiv191212699B} will be  a further step ahead. A  functional description of the evolution of $B_f$ that is more sophisticated than the simple power law assumed here is a further improvement to pursue, even though it would likely be model dependent -- see Figure~\ref{fig:Bf_sim}. A better separation of the IGM from the local origin component is desirable, to improve the estimate of the IGM RRM rms, which can be done with component separation Bayesian algorithms \citep{2016A&A...591A..13V}.

%%%%%%%%%
\section*{Acknowledgements}

We thank an anonymous reviewer for helpful comments that helped us improve the paper. 
This work has been conducted   within the LOFAR Magnetism Key Science Project\footnote{https://lofar-mksp.org/} (MKSP). 
This work has made use of LoTSS DR2 data \citep{2022A&A...659A...1S}. 
EC and VV acknowledge this work has been conducted within the INAF program METEORA. 
VV acknowledges support from Istituto Nazionale di Astrofisica (INAF) mainstream project “Galaxy Clusters Science with LOFAR” 1.05.01.86.05.
FV acknowledges financial support from the H2020 StG MAGCOW (714196).
AB acknowledges support from ERC Stg DRANOEL n.~714245 and MIUR FARE grant "SMS". 
In this work we used the {\enzo} code (\hyperlink{http://enzo-project.org}{http://enzo-project.org}), the product of a collaborative effort of scientists at many universities and national laboratories.  Our simulations were run on the  JUWELS cluster at Juelich Superc omputing Centre (JSC), under project "radgalicm2", and on the Piz Daint supercomputer at CSCS-ETHZ (Lugano, Switzerland), under project s1096, in both cases as FV as a Principal Investigator.
LOFAR \citep{2013A&A...556A...2V} is the Low Frequency Array designed and constructed by ASTRON. It has observing, data processing, and data storage facilities in several countries, which are owned by various parties (each with their own funding sources), and which are collectively operated by the ILT foundation under a joint scientific policy. The ILT resources have benefited from the following recent major funding sources: CNRS-INSU, Observatoire de Paris and Universit\'e d'Orl\'eans, France; BMBF, MIWF-NRW, MPG, Germany; Science Foundation Ireland (SFI), Department of Business, Enterprise and Innovation (DBEI), Ireland; NWO, The Netherlands; The Science and Technology Facilities Council, UK; Ministry of Science and Higher Education, Poland; The Istituto Nazionale di Astrofisica (INAF), Italy.   LoTSS made use of the Dutch national e-infrastructure with support of the SURF Cooperative (e-infra 180169) and the LOFAR e-infra group. The J\"ulich LOFAR Long Term Archive and the German LOFAR network are both coordinated and operated by the J\"ulich Supercomputing Centre (JSC), and computing resources on the supercomputer JUWELS at JSC were provided by the Gauss Centre for Supercomputing e.V. (grant CHTB00) through the John von Neumann Institute for Computing (NIC).
LoTSS made use of the University of Hertfordshire high-performance computing facility and the LOFAR-UK computing facility located at the University of Hertfordshire and supported by STFC [ST/P000096/1], and of the Italian LOFAR IT computing infrastructure supported and operated by INAF, and by the Physics Department of Turin university (under an agreement with Consorzio Interuniversitario per la Fisica Spaziale) at the C3S Supercomputing Centre, Italy.
This work made use  of the Python packages NumPy \citep{2020Natur.585..357H}, Astropy \citep{2013A&A...558A..33A}, Matplotlib \citep{2007CSE.....9...90H}, and  EMCEE \citep{2013PASP..125..306F}. Some of the results in this paper have been derived using the healpy \citep{2019JOSS....4.1298Z} and HEALPix\footnote{http://healpix.sf.net} \citep{2005ApJ...622..759G} packages. 
%%%%%%%%%%%%%%%%%%%%%%%%%%%%%%%%%%%%%%%%%%%%%%%%%%
\section*{Data Availability}

The LoTSS DR2 RM catalogue  will be publicly released as the  paper describing it (O'Sullivan et al. submitted) will be accepted for publication. The cosmological simulations of this project have been produced with the public code \small{ENZO}\footnote{enzo-project.org}. 
 Examples of the simulated datasets can be found at this URL: https://cosmosimfrazza.myfreesites.net/scenarios-for-magnetogenesis.



\bibliographystyle{mnras}
\bibliography{bevo2}

%%%%%%%%%%%%%%%%%%%%%%%%%%%%%%%%%%%%%%%%%%%%%%%%%%

%%%%%%%%%%%%%%%%% APPENDICES %%%%%%%%%%%%%%%%%%%%%

\appendix

\section{Relation between 2D and 3D probabilities of finding a source closer than a distance}\label{app:2D3D}

Assuming that a 3D probability  distribution $F$ is the same along all of the directions:  $F(x,y,z)=f((x)f(y)f(z)$, in first approximation, the 3D-probabiliity that the variable is within a distance $d$ from the centre on each direction is: 
\begin{eqnarray}
    p_{3D}(d) &=& \int_{-d}^d f(x)dx \int_{-d}^d f(y)dy \int_{-d}^d f(z)dz \\ \\
             &=& p_{1D}^3(d)\\
\textrm{where}\nonumber\\
     p_{1D}(d) &=&\int_{-d}^d f(z)dz
\end{eqnarray}
This is only a first approximation because the integration should not be extended out to $d$ in all directions. The correct $p_{3D}$ is thus smaller. 

Marginalising in  the direction $z$, we obtain the 2D-probability that the variable is within a distance $d$  along two directions only: 

\begin{eqnarray}
    p_{2D}(d) &=& \int_{-d}^d f(x)dx \int_{-d}^d f(y)dy \int_{-\infty}^\infty f(z)dz \\ \\
             &=& p_{1D}^2(d)
\end{eqnarray}
This also is  an approximation. 

Hence, 
\begin{equation}
      p_{3D}(d) \approx  p_{2D}^{3/2}(d)
\end{equation}

%%%
\section{Best fit of  RRM  assuming the magnetic field frozen to the plasma }\label{app:bestfit}
\begin{figure*}
   \centering
    \includegraphics[width=\textwidth]{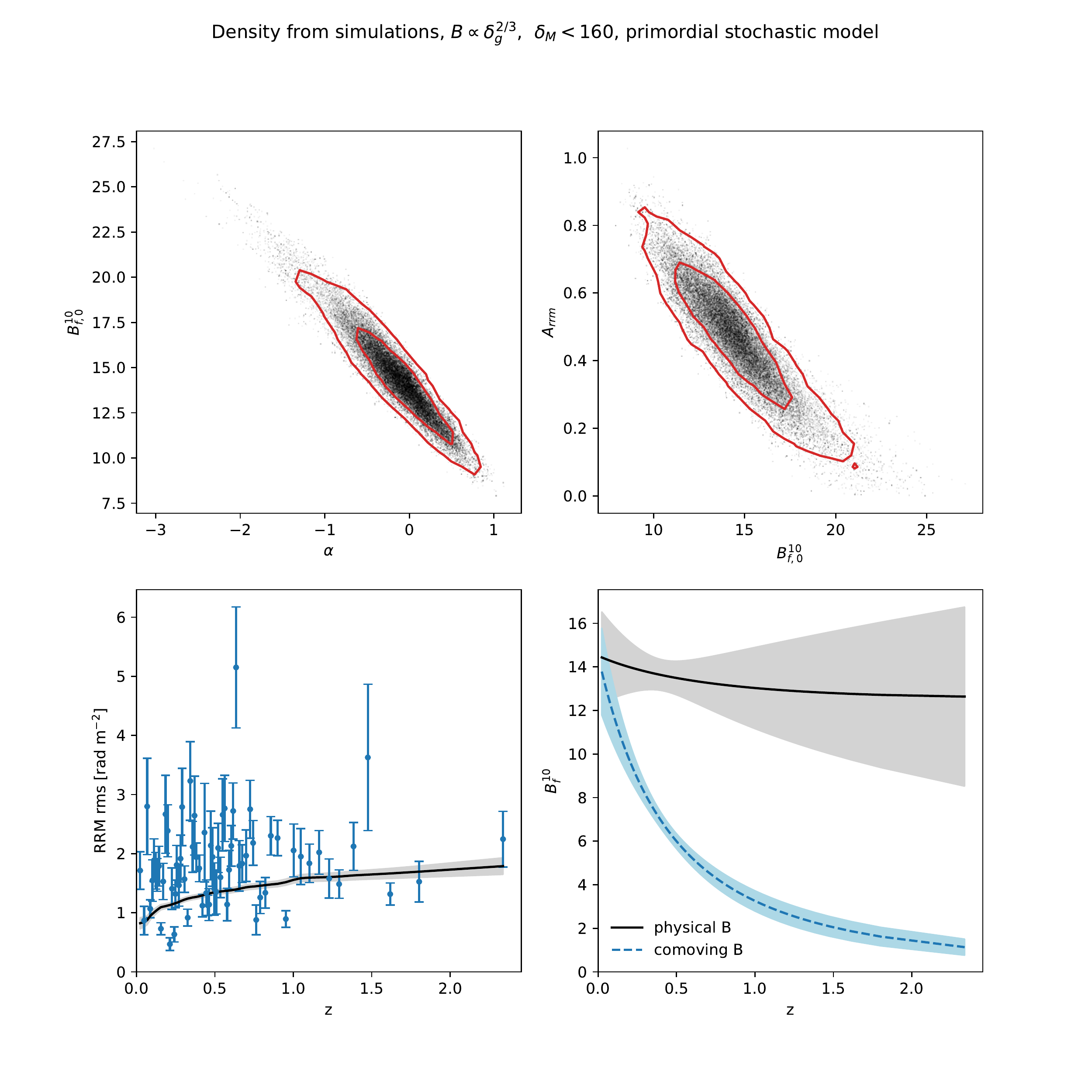}
   \caption{Best-fit results of Equation~(\ref{eq:rm_rms_a_rrmf}) to the RRM rms measured with redshift bins of 15 sources each. The filament  RRM$_f$ is computed from Equation~(\ref{eq:rm}), where the density is  taken from  cosmological simulations and the magnetic field strength depends on    the gas density as $B_f \propto \rho_g^{2/3}$.  The overdensity limit assumed is $\delta_M < 160$. The case of the primordial stochastic cosmological MHD model  is shown here. The description of the panels is as for Figure~\ref{fig:rrmfit_15}.  }
              \label{fig:rrmfit_sim}%
    \end{figure*}

%%%
Best-fit results of Equation~(\ref{eq:rm_rms_a_rrmf}) to the measured RRM rms, for the case in which   the magnetic field is assumed to be frozen to the plasma, are shown in Figure~\ref{fig:rrmfit_sim}. Only the case of the primordial stochastic scenario  is shown.

 %%%% Figure %%%%
%%%%%
%%%%%%%%%%%%%%%%%%%%%%%%%%%%%%%%%%%%%%%%%%%%%%%%%%

% Don't change these lines
\bsp	% typesetting comment
\label{lastpage}
\end{document}